\documentclass[journal]{IEEEtran}
\usepackage[utf8]{inputenc}
\usepackage{graphicx}
\usepackage{mathtools}
\usepackage{adjustbox}
\usepackage{booktabs}
\usepackage{multirow}
\usepackage{cite}
\usepackage{enumitem}

\usepackage{amsthm}
\theoremstyle{definition} % bold title, normal text

\theoremstyle{plain} % bold title, italic text

\theoremstyle{remark} % italic title, normal text
\newtheorem{remark}{Remark}

\usepackage{hyperref}
\hypersetup{
    bookmarksopen=true,
    bookmarksnumbered=true,
    breaklinks=true,
    hidelinks=true,
}

\usepackage[acronym,nomain,nopostdot]{glossaries}
\newacronym{ac}{AC}{alternating current}
\newacronym{acer}{ACER}{European Agency for the Cooperation of Energy Regulators}
\newacronym{admm}{ADMM}{Alternating Direction Method of Multipliers}
\newacronym{anova}{ANOVA}{analysis of variance}
\newacronym[firstplural={available transfer capacities (ATCs)}]{atc}{ATC}{available transfer capacity}
\newacronym{bmwi}{BMWi}{German Federal Ministry of Economic Affairs and Energy}
\newacronym{bnetza}{BNetzA}{German Federal Network Agency}
\newacronym{ca}{CA}{capacity allocation}
\newacronym{cbco}{CBCO}{critical branch under critical outage}
\newacronym{cb}{CB}{critical branch}
\newacronym{co}{CO}{critical outage}
\newacronym{cacm}{CACM}{capacity allocation and congestion management}
\newacronym{cee}{CEE}{Central Eastern Europe}
\newacronym{cm}{CM}{congestion management}
\newacronym{cne}{CNE}{critical network element}
\newacronym[firstplural=critical network elements and contingencies (CNECs)]{cnec}{CNEC}{critical network element and contingency}
\newacronym{co2}{CO\textsubscript{2}}{carbon dioxide}
\newacronym{core}{CORE}{CORE region}
\newacronym[firstplural=Central Western European (CWE)]{cwe}{CWE}{Central Western Europe}
\newacronym{d2cf}{D2CF}{2-days ahead congestion forecast}
\newacronym{da}{DA}{day-ahead}
\newacronym{dcpf}{DCPF}{linear power flow}
\newacronym{dsk}{DSK}{demand shift key}
\newacronym{der}{DER}{distributed energy resource}
\newacronym{ec}{EC}{European Commission}
\newacronym{ed}{ED}{economic dispatch}
\newacronym{eeg}{EEG}{Erneuerbare-Energien-Gesetz}
\newacronym{eex}{EEX}{European Energy Exchange}
\newacronym[firstplural=Energiewirtschaftsgesetzes (EnWG)]{enwg}{EnWG}{Energiewirtschaftsgesetz}
\newacronym{eu}{EU}{European Union}
\newacronym{euphemia}{EUPHEMIA}{Pan-European Hybrid Electricity Market Integration Algorithm}
\newacronym{fav}{FAV}{final adjustment value}
\newacronym{fba}{FBA}{flow-based allocation}
\newacronym{fbmc}{FBMC}{flow-based market coupling}
\newacronym{flh}{FLH}{full load hour}
\newacronym{frm}{FRM}{flow reliability margin}
\newacronym{gsk}{GSK}{generation shift key}
\newacronym[\glsshortpluralkey={GW}]{gw}{GW}{gigawatt}
\newacronym[\glsshortpluralkey={GWh}]{gwh}{GWh}{gigawatt hours}
\newacronym{hev}{HEV}{hybrid electric vehicle}
\newacronym{hvdc}{HVDC}{high-voltage direct-current}
\newacronym{iem}{IEM}{internal energy market}
\newacronym{jump}{JuMP}{Julia for Mathematical Programming}
\newacronym{jao}{JAO}{Joint Allocation Office}
\newacronym[\glsshortpluralkey={kV}]{kv}{kV}{kilovolt}
\newacronym[\glsshortpluralkey={kWh}]{kwh}{kWh}{kilowatt hour}
\newacronym{lp}{LP}{linear program}
\newacronym{lodf}{LODF}{load outage distribution factor}
\newacronym{lsk}{LSK}{load shift key}
\newacronym{isk}{ISK}{injection shift key}
\newacronym{mio}{mio.}{million}
\newacronym{minram}{minRAM}{minimum remaining available margin}
\newacronym[\glsshortpluralkey={Mt}]{mt}{Mt}{megaton}
\newacronym{mw}{MW}{megawatt}
\newacronym[\glsshortpluralkey={MWh}]{mwh}{MWh}{megawatt hour}
\newacronym{ndp}{NDP}{network development plan}
\newacronym[firstplural=net transfer capacities (NTCs)]{ntc}{NTC}{net transfer capacity}
\newacronym{nuts2}{NUTS2}{Nomenclature of Territorial Units for Statistics, Level~2}
\newacronym{nuts3}{NUTS3}{Nomenclature of Territorial Units for Statistics, Level~3}
\newacronym{opsd}{OPSD}{Open Power System Data}
\newacronym{or}{OR}{Operations Research}
\newacronym{opf}{OPF}{optimal power flow}
\newacronym{otc}{OTC}{over-the-counter}
\newacronym{pcr}{PCR}{Price Coupling of Regions}
\newacronym[sort=pomato]{pomato}{\begin{small}POMATO\end{small}}{Power Market Tool}
\newacronym{psp}{PSP}{pumped-storage plants}
\newacronym{pv}{PV}{photovoltaik}
\newacronym{ptdf}{PTDF}{power transmission distribution factor}
\newacronym{px}{PX}{power exchange}
\newacronym{ram}{RAM}{remaining available margin}
\newacronym{res}{RES}{renewable energy sources}
\newacronym{scopf}{SCOPF}{security-constrained optimal power flow}
\newacronym{tso}{TSO}{transmission system operator}
\newacronym[\glsshortpluralkey={TWh}]{twh}{TWh}{terawatt hour}
\newacronym{tyndp}{TYNDP}{Ten-Year Network Development Plan}
\newacronym[sort=uebertr]{unb}{\"UNB}{Übertragungsnetzbetreiber}

\glsdisablehyper
    
\usepackage{amsmath, amssymb, amsthm, bm}
\usepackage{amsfonts}
\usepackage{xurl}
\usepackage{float} 
\usepackage{wrapfig}
\usepackage{color, soul}
\usepackage{csquotes}
\usepackage{tikz}
\usepackage{units}

\DeclareMathOperator{\PTDF}{\mathbf{PTDF}}
\DeclareMathOperator{\zPTDF}{\mathbf{zPTDF}}
\DeclareMathOperator{\FRM}{\mathbf{FRM}}
\DeclareMathOperator{\RAM}{\mathbf{RAM}}
\DeclareMathOperator{\minRAM}{minRAM}
\DeclareMathOperator{\GSK}{\mathbf{GSK}}
\DeclareMathOperator{\FAV}{\mathbf{FAV}}
\DeclareMathOperator{\EX}{EX}
\DeclareMathOperator{\EXvec}{\mathbf{EX}}
\DeclareMathOperator{\Var}{Var}
\DeclareMathOperator{\NP}{NP}
\DeclareMathOperator{\NPvec}{\mathbf{NP}}
\DeclareMathOperator{\ntc}{NTC}
\DeclareMathOperator{\sgn}{sgn}

\newcommand{\set}[1]{\mathcal{#1}} % for caligraphed set-symbols
\newenvironment{ldescription}[1]
  {\begin{list}{}%
   {\renewcommand\makelabel[1]{##1\hfill}%
   \settowidth\labelwidth{\makelabel{#1}}%
   \setlength\leftmargin{\labelwidth}
   \addtolength\leftmargin{\labelsep}}}
  {\end{list}}
  
\title{Uncertainty-Aware Capacity Allocation\\in Flow-Based Market Coupling}
\author{Richard Weinhold,~\IEEEmembership{Member,~IEEE} and Robert Mieth,~\IEEEmembership{Member,~IEEE}

\thanks{This work was partially funded by the Bundesministerium für Wirtschaft und Energie grant No.~03EI1019B.}%
\thanks{R.~Weinhold and R.~Mieth contributed equally to this work.}}

\begin{document}

\maketitle

\begin{abstract}
    The effective allocation of cross-border trading capacities is one of the central challenges for the implementation of a pan-European internal energy market. 
    In contrast to traditional power flow-ignorant methods like net transfer capacities (NTC), flow-based market coupling (FBMC) has been shown to increase price convergence between market areas, while improving congestion management effectiveness.
    However, explicitly analysing FBMC for a future power system with a very high share of intermittent renewable generation is often overlooked in the current literature. 
    This paper provides a comprehensive summary on the technical specification of the FBMC process and FBMC modeling methods. It discusses implications of policy considerations and explicitly discusses the impact of high-shares of intermittent generation on FBMC performance.
    Further, we propose probabilistic security margins compatible with the current FBMC implementation to better account for renewable uncertainty in FBMC modeling.
    We conduct numerical experiments on the IEEE 118 bus test system to showcase the proposed model formulations and our data and implementation is published open source.
\end{abstract}


\section*{Nomenclature}
\newcommand{\longestitem}{$C_{0,o,g}$}

\noindent\textit{A.Indices}
\begin{ldescription}{\longestitem}
   \item[E] Number of CNEs
   \item[L] Number of lines
   \item[N] Number of nodes
   \item[Z] Number of zones
\end{ldescription}

\noindent\textit{B. Parameters}
\begin{ldescription}{\longestitem}
    \item [$\bm{d}_t$]  Vector of demand at time $t$ indexed as $d_{t,i}$
    \item [$\bm{e}$]  Vector of ones in appropriate dimensions
    \item [$\bm{f}^{ref}$] Vector of reference flows computed from basecase and day-ahead results
    \item [$\overline{\bm{f}}$] Vector of maximum line capacity indexed by $\overline{f}_j$
    \item [$\FAV$] Vector of final adjustment values for each CNEC
    \item [$\FRM$] Vector of flow reliability margins for each CNEC
    \item [$\overline{\bm{g}}$] Vector of maximum generation capacity indexed by $\overline{g}_i$
    \item [$\GSK$] Vector of generation shift keys
    \item [$\minRAM$] Minimum remaining available margin 
    \item [$\ntc_{k,k'}$] Net transfer capacity between zone $k$ and $k'$
    \item [$\PTDF$] Nodal power transfer distribution matrix
    \item [$\bm{r}_t$]  Vector of available RES power at time $t$ indexed as $r_{t,i}$
    \item [$\bm{R}$]  Vector of linear cost parameters
    \item [$\RAM$] Vector of remaining available margins on each CNEC
    \item [$s_t$] Square root of sum of covariance matrix $s^2 = \bm{e}^T\bm{\Sigma_t}\bm{e}^T$
    \item [$z_\epsilon$] Risk parameter defined as $\Phi^{-1}(1-\epsilon)$
    \item [$\zPTDF_t$] Zonal power transfer distribution matrix at time $t$ where $\zPTDF_{t,j}$ is the $j$-th row of $\zPTDF_t$
    \item [$\epsilon$] Risk level
    \item [$\bm{\Sigma}_t$] Covariance matrix of forecast error at time $t$
    \item [$\bm{\omega}_t$] Forecast error at time $t$
    \item [$\bm{\Omega}_t$] Uncertainty space of forecast error at time $t$
\end{ldescription}

\vspace{0.5em}
\noindent\textit{C. Variables}
\begin{ldescription}{\longestitem}
    \item [$\bm{C}_t$] Vector of curtailment at time $t$ indexed as $C_{t,i}$
    \item [$\EXvec_t$] Matrix collecting the net exchange between zones at time $t$
    \item [$\bm{f}_t$] Vector of flows at time $t$ indexed as $f_{t,j}$ 
    \item [$\bm{G}_t$] Vector of active generation at time $t$ indexed as $G_{t,i}$
    \item [$\bm{I}_t$] Vector of nodal net injections at time time $t$ indexed as $I_{t,i}$
    \item[$\NPvec_t$] Vector of net positions at time $t$ indexed as $\NP_{t,k}$
    \item[$T_{t,j}$] Auxiliary variable capturing the standard deviation of flow on CNE $j$ at time $t$
    \item[$\bm{\alpha}_t$] Vector of balancing participation factors at time $t$ indexed by $\alpha_{t,i}$
\end{ldescription}

\vspace{0.5em}
\noindent\textit{D. Mappings, Operators, and Other}
\begin{ldescription}{\longestitem}
    \item[$\cdot^{bc}$] Value obtained from basecase stage
    \item[$\mathcal{C}(\cdot)$] Generator cost model
    \item[$\cdot^{da}$] Value obtained from day-ahead stage
    \item[$\mathbb{E}(\cdot)$] Expected value
    \item [$\mathcal{F}^\text{nodal}$] Space of feasible net injections for nodal network model
    \item [$\mathcal{F}^\text{ntc}$] Space of feasible net exchanges for NTC model
    \item [$\mathcal{F}^\text{zonal}$] Space of feasible net positions for zonal network model
    \item[$\mathcal{L}(i)$] Returns line index of CNEC $i$
    \item[$\bm{m}$] Maps nodes to zones
    \item[$\mathcal{P}(\cdot)$] Curtailment penalty model
    \item[$\mathbb{P}(\cdot)$] Probability
    \item[$\cdot^{red}$] Value obtained from redispatch stage
    \item[$\Var(\cdot)$] Variance
    \item[$\sigma(\cdot)$] Standard deviation
    \item [$\Phi$] Cumulative distribution function of the standard normal distribution
    \item [$\lVert \cdot \rVert_2$] 2-norm
\end{ldescription}

\section{Introduction}
\glsresetall

In the interconnected European power system, transnational electricity trading promises to improve system efficiency, market liquidity and price convergence \cite{schonheit_fundamental_2021}. 
However, the transport capacity between national market areas is limited and depends on power flow physics. 
\Gls{fbmc} has been introduced as a method to compute these capacities such that cross-border trade is as unconstrained as possible, while ensuring system security and operational efficiency.
\gls{fbmc} addresses many deficiencies of previous capacity allocation methods (e.g., net transfer capacities) by modeling the relationship between forecast market outcomes, the expected generator dispatch supporting this market outcome, and the resulting physical power flow through critical transmission equipment.
However, the models and forecasts underlying the \gls{fbmc} process are, by nature, imperfect. Additional real-time congestion management, e.g., out-of-market generator redispatch, is required to continuously ensure system security.
These real-time corrections are a central aspect of the overall economic evaluation of \gls{fbmc} and their importance is amplified in a power system with predominantly intermittent renewable power production and declining dispatchable generation capacities.
Hence, while the benefits of \gls{fbmc} have been convincingly demonstrated for today's portfolio of generation assets, its effectiveness for future renewable-dominant power system is subject of an ongoing debate among market participants, regulators and research scholars. 
This paper contributes to the currently scarce literature on \gls{fbmc} in a highly renewable power system as follows. 
First, we describe a fundamental model of \gls{fbmc} based on publicly available documentation. Here, we clearly distinguish between modeling choices that define the parameters of the \gls{fbmc} process and the required simulation of the market-clearing and real-time dispatch processes.
Second, we propose a new approach that internalizes statistical information of renewable generation forecasts into the \gls{fbmc} routine. This allows to substitute ad-hoc fixed security margins for a higher fidelity model that more closely matches current system operator practices and requirements. 
Third, we demonstrate our models and implementation on the IEEE 118 bus test system and discuss \gls{fbmc} efficiency in comparison to two benchmarks. 

\subsection{Background and related literature}

The European electricity system is a collection of national zonal electricity markets.
The development of the European \gls{iem} aims to establish a pan-European electricity trading platform under a policy triangle of security of supply, affordability, and sustainability \cite{europeancommission_directive_1997}. 
Organizing the vast majority of Europe's electricity consumption in a coordinated market framework is the achievement of over 20 years of continuous policy iteration for more effective market designs and procedures \cite{glachant_achievement_2010}.
Central to these market coupling procedures are capacity allocation and congestion management routines, i.e., methods to determine exchange capacities between market areas such that transnational trade volumes are high, while secure system operation is guaranteed.
In the early stages of the \gls{iem}, the development of efficient methods for capacity allocation and congestion management were motivated by the scarcity of transmission capacity inherited from pre-liberalized market structures \cite{meeus_electricity_2008}.
Current developments, on the other hand, are mostly driven by the system's transformation towards a sustainable energy system \cite{directorategeneralforenergy_clean_2019}. 
Specifically, in the context of a generation portfolio that is characterized by more intermittent renewable energy sources and fewer dispatchable generators, increasing cross-border cooperation promises a more efficient utilization of available infrastructure by improving the coordination of (flexible) generation resources. 

To date, various market coupling methods are in use.
While most neighboring market zones rely on \glspl{ntc} or \glspl{atc}, \gls{cwe}, comprising France, Belgium, Luxembourg, Netherlands, Germany and Austria, has adopted the target method \gls{fbmc}.

\gls{fbmc} differs from \gls{ntc} and \gls{atc} in how transmission capacity is allocated to markets. 
\gls{ntc} and \gls{atc} are independent limits on the allowable exchange between two adjacent market areas and reflect the physical capacity of the transmission lines between those areas minus some security margins.
This approach, however, ignores interdependencies of cross-border flows caused by the physics of power flow in an interconnected network.
\gls{fbmc}, on the other hand, derives limits on the net exports (so called ``net positions'') of all involved bidding zones \textit{simultaneously} and, thus, can capture interdependencies between trading activities and power flows more holistically for the entire coupled market.

As a result, physical transmission constraints imposed by limited thermal line capacities are better reflected in the market clearing solution, while, at the same time, high cross-border trade capacity is ensured. 
This promises more efficient capacity allocation and improved inter-zonal price convergence~\cite{schonheit_fundamental_2021}.

Although \gls{fbmc} was inaugurated only in 2015 as part of EC Directive 15/1222 \cite{europeancommission_commission_2015}, it was conceived and designed already a decade earlier \cite{etso_coordinated_2001}. 
As a result, \gls{fbmc} was motivated by necessary improvements to capacity allocation in the \gls{iem} and not by a need to realize policy targets, e.g., accommodate high shares of \gls{res}.
Yet, policy specifications are indeed applied to \gls{fbmc} regulations. 
For example, as part of the Clean Energy Package \cite{directorategeneralforenergy_clean_2019} and the accompanying update to the regulation on capacity allocation and congestion management, the 2015 EU Commission regulation \cite{europeancommission_commission_2019} requires \glspl{tso} to allocate a minimal share of physical line capacity to the market. 
This kind of engagement with specifics of the process illustrates the regulatory willingness to closer engage with market design to align the process with sustainability targets. Therefore, \gls{fbmc}, as an important part of the \gls{iem}, should become an active element of the transformation process towards a decarbonized energy system.  

For any policy developments, quantitative analyses of the \gls{fbmc} process are an important basis, as they provide reliable context for necessary design decisions.
Initial publications on \gls{fbmc} stem from conceptional documentations of the involved \glspl{tso} and power exchanges \cite{etso_coordinated_2001, etso_flowbased_2004}. As part of a ``dry-run'', i.e., simulated operation, in 2008 \cite{amprion_cwe_2011} and a parallel run in 2013 \cite{rte_cwe_2015} the process was evaluated and its effectiveness in achieving higher welfare and improving price convergence was verified.
Additionally, \gls{fbmc} is described in a continuously updated documentation by the involved \glspl{tso} \cite{50hertz_documentation_2020}. 
Complementary descriptions and research of the involved parameters was published in different articles from \gls{tso} work-groups \cite{schavemaker_flowbased_2008, aguado_flowbased_2012, marien_importance_2013}. 

Independent academic publications on \gls{fbmc} have mainly focused on the question of how the process can be modeled accurately and how model assumptions and parameter choices affect market outcomes and system operations. 
For example, Byers and Hug \cite{byers_modeling_2020} provide insights on the fundamental \gls{fbmc} modeling process and perform a numerical sensitivity analyses for the various modeling parameters.
Schönheit \textit{et al.} (2020) \cite{schonheit_impact_2020} and Finck \textit{et al.}\cite{finck_impact_2018} evaluate the impact of different generation shift keys, a core parameter that captures the participation of generators in the zonal market outcome and, thus, maps market outcomes to power flows on selected network elements.
Schönheit \textit{et al.} (2021a) \cite{schonheit_improved_2021} discuss the selection of these network elements in the context of market zone configurations.

\subsection{Motivation and Contribution}

The above mentioned publications \cite{byers_modeling_2020,schonheit_impact_2020,finck_impact_2018,schonheit_improved_2021} focus on specific modeling and parameter choices and present FBMC as a pure technical process.
However, we argue that \gls{fbmc} is closely related to policy-making and has the ability to accommodate policy goals, e.g, enforcing minimal trading capacities or facilitating \gls{res} expansion.

Schönheit \textit{et al.}~\cite{schonheit_fundamental_2021} touch on this topic by describing the \glspl{tso}' degree of freedom to influence \gls{fbmc} parametrization and influence market results. 
The analysis also includes a scenario with a moderate increase in \gls{res} capacities on a synthetic test system. 
Matthes \textit{et al.}~\cite{matthes_impact_2019} analyse \gls{fbmc} in for the target year 2025 with higher shares of \gls{res}, but the implications on \gls{fbmc} are not explicitly discussed. 
However, the increasingly renewable-dominated generation mix requires such a discussion because \gls{res} uncertainty directly challenges the fundamental assumption of \gls{fbmc} that \glspl{tso} can forecast the grid situation at point-of-dispatch accurately. 
As a result, reliability margins, which are modelled as static values in the existing literature \cite{schonheit_fundamental_2021,byers_modeling_2020,schonheit_impact_2020,finck_impact_2018,schonheit_improved_2021,matthes_impact_2019}, become more relevant in their role of robustifying the market outcome against forecast errors. 

Motivated by the literature gap outlined above, this paper makes the following contributions. 
\begin{enumerate}
    \item We derive the fundamental relationship between, \gls{fbmc} parameters and market outcomes, highlight how \gls{fbmc} can accommodate policy goals, and discuss central modeling choices in the context of regulatory requirements.
    \item 
    We propose an improved approach to internalize risk from forecast uncertainty into the \gls{fbmc} parameter computation process.
    This approach is in line with risk evaluation techniques currently employed by \glspl{tso} and resembles an already established linear generator control model. We show that the proposed approach produces more efficient security margins on critical network elements and improves \gls{fbmc} efficiency for renewable-dominant electricity markets.
    \item We evaluate the effectiveness of \gls{fbmc} with and without the proposed uncertainty-aware modification by benchmarking it against capacity allocation via \gls{ntc}/\gls{atc}. All data and code is published and available open-source.
\end{enumerate}

\section{FBMC Concept and Simulation}\label{sec:nrel-fb_concept}

\begin{figure}
    \centering
    \includegraphics[width=0.7\columnwidth]{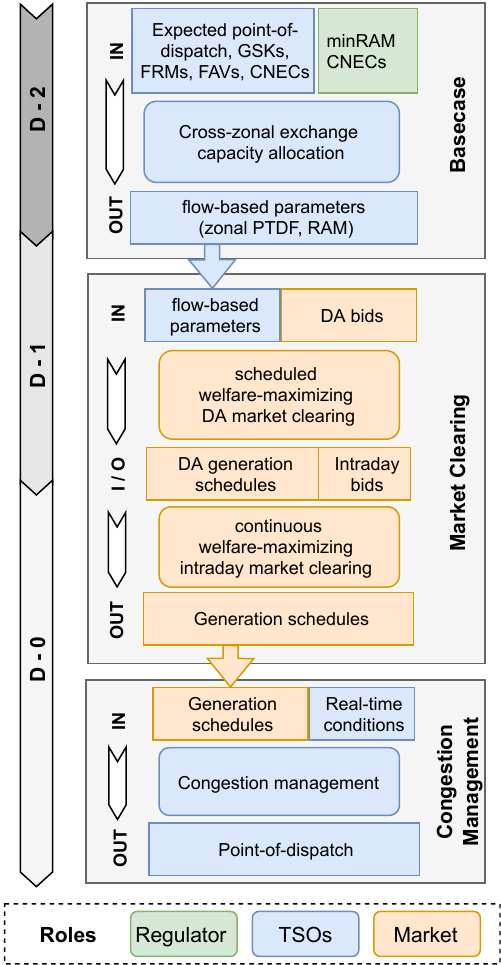}
    \caption{FBMC process overview.}
    \label{fig:nrel-fb_process}
\end{figure}

\gls{fbmc} is a three-stage process designed to allocate commercial exchange capacity between adjacent electricity markets (i.e., cross-border exchange). 
It is coordinated by the local \glspl{tso} and power exchanges and aims to accommodate inter-zonal electricity trading with respect to available transmission system capacity and physical power flows.
Fig.~\ref{fig:nrel-fb_process} provides an overview of the three \gls{fbmc} stages and shows their timing, involved stakeholders, parameters and results. 
For any given day $D$, the \gls{fbmc} process starts two days in advance at $D$--2, with the calculation of the so called \textit{basecase}.
The basecase is informed by \gls{tso}-generated forecasts on the expected point-of-dispatch and provides ``\textit{a best estimate of the state of the [...] system for day D}'' \cite[p. 26]{50hertz_documentation_2020}.
Based on these forecasts, on additional predefined policies, and on regulatory constrains (see discussion {below}), the \glspl{tso} calculate the so called \textit{flow-based parameters} that are used to constrain the commercial exchange in the following \textit{market clearing} stage. 
Generation and load bids collected from all bidding areas that are part of the \gls{cwe} region are cleared in the day-ahead (DA) and intraday markets at $D$--1 and $D$, respectively.
Lastly, during the \textit{re-dispatch} stage, the \glspl{tso} may require changes to the generators' final point-of-dispatch to resolve any network congestion or other threats to system security due to real-time conditions (e.g., load and \gls{res} forecast deviations).

After clarifying some preliminaries, we will first derive and clarify the definitions and parameters of the flow-based formalism as closely to the regulatory language as possible. Next, we present a model formulation to synthesize the three-step \gls{fbmc} process. Finally, we augment this model to internalize forecast error statistics to choose line reliability margins more effectively.

\subsection{Preliminaries}\label{sec:nrel-formal_fbmc}

The \gls{fbmc} process and the definition of the flow-based parameters relies on a model of the physical transmission system, which we formalize as follows. 
Consider an interconnected transmission network with $N$ nodes and $L$ lines. Further, let each node $i$ be defined by its net power injection $I_i$ collected in column vector  $\bm{I}=[I_i]_{i=1}^N\in\mathbb{R}^N$. If $I_i>0$, then node $i$ is a net generator, if $I_i<0$, then node $i$ is a net load.
For each vector of net injections $\bm{I}$ there exists a vector $\bm{f}=[f_j]_{j=1}^L\in\mathbb{R}^L$ that collects the flows along each line $j$ and is defined by the physics of power flow in a network. 
As derived in detail in, e.g., \cite{weinhold_fast_2020}, power flow physics in high-voltage transmission networks allow a linear approximation of the relationship between $\bm{f}$ and $\bm{I}$ using a \textit{power transfer distribution factor} matrix $\PTDF\in\mathbb{R}^{L\times N}$ such that 
\begin{equation}
    \bm{f} = \PTDF \cdot \bm{I}.
\end{equation}
The maximum allowable flow (capacity) of each line $j$ is given by $\overline{f}_j$ collected in vector $\overline{\bm{f}} = [\overline{f}_j]_{j=1}^L$.

Finally, all nodes of the network are grouped in $Z$ compact and disjoint \textit{market zones}.
The sum of the injections of all nodes in one zone is called the \textit{net position} $\NP_k$ of zone $k$. If $\NP_k>0$, then zone $k$ is a net exporter and if $\NP_k<0$, then zone $k$ is a net importer. All net positions are collected in vector $\NPvec = [\NP_k]_{k=1}^Z\in\mathbb{R}^Z$. 

\subsection{Flow-based formalism}

The effectiveness of \gls{fbmc} to enable least-cost, yet physically feasible day-ahead market outcomes across interconnected market zones hinges on the precise definition of the flow-based parameters, which specify (i) how cross-border power exchange affects power flow on transmission lines and (ii) how much capacity on each line is available to accommodate flow caused by cross-border exchange. 
Each specification (i) and (ii) is given by the following respective parameters:
\begin{enumerate}[label=(\roman*)]
    \item \textbf{Zonal PTDF}; Matrix $\zPTDF\in\mathbb{R}^{E \times Z}$ maps net position vector $\NPvec$ to the flow on a \textit{selection} of lines in the network, the so called \textit{critical network elements} (CNE). \glsunset{cne}
    \glspl{cne} can be selected to capture flows in either direction of a specific line. As a result, the number of CNEs $E$ is in the interval $[1,2L]$.
    \item \textbf{\glsreset{ram}\Gls{ram}}; For each \gls{cne}, vector $\RAM\in\mathbb{R}^{E}$ defines the capacity in the direction of each \gls{cne} that is available for cross-border trading in the day-ahead market.
\end{enumerate}
See also the top box of Fig.~\ref{fig:nrel-fb_process}.
Although these parameters are generated, exchanged, and published following specific rules requested by regulation and laid out in the documentation of the \gls{fbmc} process in \cite{50hertz_documentation_2020}, they depend on the exact choice of the underlying forecast and meta-parameters that are either chosen at the discretion of the \glspl{tso} or given by policy requirements. 

The exact definition of $\zPTDF$ depends on the chosen \glspl{cne}, additional contingency scenarios, and so called \textit{\glspl{gsk}}.

\subsubsection{CNE/CNEC}
\label{sssec:cne_cnec}

``\textit{A CNE is considered to be significantly impacted by CWE cross-border trade, if its maximum CWE zone-to-zone PTDF is larger than a threshold value that is currently set at 5\%.}'' 
\hspace*{\fill}\cite[p. 19f]{50hertz_documentation_2020}

Network elements are considered \gls{cne} only if \glspl{tso} determine that their flow is significantly driven by cross-zonal exchange. This selection prevents heavily loaded lines that are largely insensitive to changes in the zones' net positions to limit inter-zonal exchange capacity allocation.
Further, for each network element the \glspl{tso} identify a set of contingencies (C), i.e., unplanned outages of network elements, that highly impact the flow on said \gls{cne}. 
The resulting set of \glspl{cnec}, the full $\PTDF$ matrix and additional linear outage factors, which capture flow shifts during the selected contingencies, form the basis for $\zPTDF$ computation. 

\subsubsection{GSKs}
``\textit{A GSK aims to deliver the best forecast of the impact on Critical Network Elements of a net position change [and] is calculated according to the reported available market driven power plant potential of each TSO divided by the sum of market driven power plant potential in the bidding zone.}''

\hspace*{\fill}\cite[p. 38ff]{50hertz_documentation_2020}

Since line flows $f$ depend on the injection at every node, the computation of $\zPTDF$ requires an estimation of how changes in net position $\NPvec$ are distributed to changes in net injections $I$. 

Assuming that the difference between forecasted net position in the basecase and the realized net position in the market phase is small, the set of generators that will serve this difference by ``shifting'' their production levels can be anticipated. 
The resulting nodal contribution to net position changes is formally captured in matrix $\GSK\in\mathbb{R}^{N\times Z}$ that maps a change $\Delta\!\NPvec$ in net positions to a change in nodal injections $\Delta \bm{I}$ such that 
\begin{equation}
    \Delta \bm{I} = \GSK \cdot \Delta\!\NPvec.
    \label{eq:gsks}
\end{equation}
Eq.~\eqref{eq:gsks} implies an alternative definition of $\zPTDF$ as
\begin{equation}
    \zPTDF =  \begin{bmatrix}
        \sgn_{\mathcal{L}(1)} \PTDF_{\mathcal{L}(1)} \\
        \vdots \\
        \sgn_{\mathcal{L}(E)} \PTDF_{\mathcal{L}(E)}
    \end{bmatrix} \cdot \GSK,
\end{equation}    
where $\PTDF_{\mathcal{L}(i)}$ is the row of the PTDF matrix corresponding to the $i$-th CNE and $\sgn_{\mathcal{L}(i)}$ indicates whether CNE $i$ is defined as a positive or negative flow.
Note that GSKs implicitly attribute all possible changes in net position $\NPvec$ to changes of dispatchable generators and is ignorant to any changes in $\NPvec$ caused by forecast errors of load and \gls{res}. 
Next, the final \gls{ram} values are corrected by additional \textit{flow reliability margins} (FRMs) and \textit{final adjustment values} (FAVs). \glsunset{frm}\glsunset{fav}

\subsubsection{FRM}
``\textit{[F]or each Critical Network Element, a Flow Reliability Margin (FRM) has to be defined, that quantifies at least how [...] uncertainty impacts the flow on the Critical Network Element.}'' 
\hspace*{\fill}\cite[p. 47]{50hertz_documentation_2020}

\glspl{frm} are static \gls{ram} reductions that are informed by past observations on how the flow on each \gls{cne} changed between the basecase forecast and the point-of-dispatch realization. 
Therefore, \gls{frm} captures load and \gls{res} forecast uncertainty.

\subsubsection{FAV}
``\textit{With the Final Adjustment Value (FAV), operational skills and experience [...] can find a way into the Flow Based-approach by increasing or decreasing the remaining available margin (RAM) on a CNE for very specific reasons [...] to eliminate the risk of overload on the particular CNE.}''
\hspace*{\fill}\cite[p. 25]{50hertz_documentation_2020}

\glspl{fav} decrease or increase \gls{ram} based on \glspl{tso} operational experience. It may reflect remedial actions at the point-of-dispatch or other complex system security considerations.

Finally, the available commercial exchange capacity between market zones is defined by constraining the zonal net positions \textit{relative to the basecase}.
For the basecase, the \glspl{tso} forecast the expected flow $\bm{f}^{bc}$ on all \glspl{cne} and the expected net positions $\NPvec^{bc}$. The net positions that can be realized during the market stage ($\NPvec^{da}$) can only differ from $\NPvec^{bc}$ if \gls{cne} limits $\overline{\bm{f}}$ (corrected by \glspl{frm} and \glspl{fav}) are maintained:
\begin{subequations}\label{eq:nrel-fb_parameters}
    \begin{align}
        \zPTDF (\NPvec^{da} - \NPvec^{bc}) &\leq  \overline{\bm{f}} - (\FRM + \FAV) - \bm{f}^{bc}.  \label{eq:nrel-fb_parameters_1}
    \end{align}

    Note that the explicit introduction of $\NPvec^{bc}$ in \eqref{eq:nrel-fb_parameters_1} is necessary to ensure the validity of the \glspl{gsk} and, thus, $\zPTDF$. See also \cite{schonheit_impact_2020} for a broader discussion.

    Eq.~\eqref{eq:nrel-fb_parameters_1} can be rewritten as follows:
    \begin{align}
        &\zPTDF \cdot \NPvec^{da} \leq \overline{f} -\!(\FRM + \FAV)\!-\!\bm{f}^{bc} && \\
        & \hspace{4cm} + \zPTDF \cdot \NPvec^{bc} \nonumber &&\\
        \Leftrightarrow\ &\zPTDF \cdot \NPvec^{da} \leq \overline{f} -\!(\FRM + \FAV)\!-\!\bm{f}^{ref} \label{eq:nrel-fb_parameters_3} &&\\
        \Leftrightarrow\ \Aboxed{&\zPTDF \cdot \NPvec^{da} \leq \RAM}.&& \label{eq:nrel-fb_parameters_4}
    \end{align}
\end{subequations}

In Eq.~\eqref{eq:nrel-fb_parameters_3}, $\bm{f}^{ref} = \bm{f}^{bc} - \zPTDF \cdot \NPvec^{bc}$ denotes the \textit{reference flow} that captures a residual between the parameter choices made in $\zPTDF$ and the forecasted $\bm{f}^{bc}$. While $\bm{f}^{ref}$ can be assumed small, it is not necessarily zero. 
Eq.~\eqref{eq:nrel-fb_parameters_4} yields the desired limit on market-based net positions $\NPvec^{da}$ subject to the flow-based parameters $\zPTDF$ and $\RAM$ \cite[p. 60]{50hertz_documentation_2020}. 
The space of all possible net positions that fulfill \eqref{eq:nrel-fb_parameters_4} is called \textit{flow-based domain}.
Fig.~\ref{fig:nrel-domain_with_frm} shows an example for a flow-based domain between two zones.

Regardless of the formal \gls{ram} definition, regulations, e.g., \cite[Art. 16]{europeancommission_commission_2019}, prescribe specific conditions that define a minimal percentage of \gls{cne} capacity that must be made available for cross-border exchange and that ensures the feasibility of long-term traded capacities called \gls{minram}.

\subsubsection{minRAM}
``\textit{CNEs with a RAM of less than the minRAM [...] are assigned an AMR value (adjustment for minRAM) in order to increase the RAM.}''
\hspace*{\fill}\cite[p. 64]{50hertz_documentation_2020}
\newline
The $\minRAM$ defines a lower bound for the RAM based on the \glspl{cne} capacity and is applied after \glspl{frm} and FAVs:
\begin{align}
\RAM = \max(\minRAM \cdot \overline{\bm{f}}, \overline{\bm{f}} - (\FRM + \FAV) - \bm{f}^{ref}).
\end{align}

\subsubsection{Long-term allocations}
``\textit{The long-term-allocated capacities of the yearly and monthly auctions have to be included in the initial Flow Based-domain}''
% \newline 
\hspace*{\fill}
\cite[p. 66]{50hertz_documentation_2020} \newline
This requirement ensures that trades on energy futures and bilateral delivery contracts outside of the day-ahead or intraday market clearing stage remain feasible within the flow-based domain.

\glspl{frm}, \glspl{fav}, \gls{minram} and long-term allocations either enlarge the flow-based domain to enable higher price convergence (\gls{minram}), shrink the flow-based domain to accommodate security margins (\gls{frm}), or go both ways (\gls{fav}). 

\subsection{Flow-based discussion}

It is clear, that the resulting flow-based parameters do not only capture formal definitions, but also internalize methods to account for uncertainty and imperfections, e.g., arising from zonal aggregation and forecasts.
Further, their specific computation significantly depends on \textit{policy} considerations that externally define the desired level of available commercial cross-border exchange.  
For example, regulation states a clear goal of achieving higher price convergence \cite{europeancommission_directive_2019}. As a result, only cross zonal tie lines are encouraged to be nominated as \glspl{cne} \cite[p.8]{acer_annual_2018} and a \gls{minram} of 70\% will be required by 2025. This indicates that, independent of the actual grid situation, regulation enforces large trading domains and that potentially higher cost for congestion management (e.g., real-time redispatch) fall into the responsibility of local \glspl{tso}. 

In previous academic studies, derivation and application of the flow-based parameters is mostly understood as a strictly formal process that only captures the physical transmission state and is largely independent of policy considerations.
These studies generally focus on the formal dimension in their numerical experiments by describing the relation of a specific parametrization policy to a chosen metric, e.g., system cost or welfare, with the goal to provide a better understanding of parameter choices.
Current literature on flow-based parameter policies in relation to system cost exist for \glspl{gsk} \cite{voswinkel_flowbased_2019}, \glspl{minram} \cite{schonheit_minimum_2021}, commercial exchange and uncertainty in the basecase parametrization \cite{byers_modeling_2020} and selection of \glspl{cnec} \cite{schonheit_fundamental_2021}.
All contribute to better understanding the relation between the parameters. However, comparability remains difficult since it requires similar definitions on how flow-based domains should be used.
Most studies do not explicitly discuss which overall target the capacity allocation strives for. 

Since the current regulation explicitly requires a parametrization to provide higher exchange capacities to the markets with the goal to ensure the integration of higher shares of \gls{res} \cite{europeancommission_commission_2019}, it is important to make these considerations part of the modeling process. Without such considerations the effectiveness of \gls{fbmc} to accommodate higher shares of \gls{res} cannot be definitively answered.
With this paper we aim to contribute to this discussion by providing a transparent parametrization of the \gls{fbmc} process and the underlying market simulations and by numerically showing the effects of higher shares of \gls{res}.
We explicitly discuss different consideration regarding the permissiveness of day-ahead trading domains by \gls{minram} and \gls{cnec} selection and the effect on total system cost and congestion management. 
In addition, we provide a sensible way to include risk-aware security margins \glspl{frm} in the modeling process. The permissive capacity allocation in systems with high shares of intermittent generation raises the question of operability. Thus, we include process-considerations regarding expected deviations from scheduled generation to make the system more robust. 

\section{Model Formulation}

\subsection{Market Simulation}
\label{sec:nrel-formulation}

In addition to computing flow-based parameters, modeling and studying the three-step \gls{fbmc} process as shown in Fig.~\ref{fig:nrel-fb_process} requires a simulation of basecase, market clearing and congestion management processes. 
We model all of these steps as a multi-period \gls{ed} problem, where each step is constrained by a specific set of network or transport constraints. 

\subsubsection{Base Model}

The \gls{ed} is given as:
\begin{subequations}\label{eqs:general_economic_dispspatch}%
    \begin{align}      
    \min_{\substack{\{\bm{G}_t, \bm{C}_t,\\\EXvec_{t}, \NPvec_{t}\}}} \quad 
       \hspace{-1cm} &\hspace{1cm} \sum_{t=1}^T [\mathcal{C}(\bm{G}_t) + \mathcal{P}(\bm{C}_t)] &&  \label{eq:nrel-obj}\\
    \text{s.t. }\quad   & \forall t=1,...,T:  && \nonumber \\
    &\bm{0} \leq \bm{G}_t \leq \overline{\bm{g}} &&  
        \label{eq:nrel-capacity_g}\\
    &\bm{0} \leq \bm{C}_t \leq \bm{r}_t &&  \label{eq:nrel-capacity_c}\\
    & \bm{G}_t + (\bm{r}_t - \bm{C}_t) - \bm{d}_t = \bm{I}_t && \label{eq:nrel-eb_nodal}\\  
    &\bm{m} (\bm{G}_t + (\bm{r}_t - \bm{C}_t) - \bm{d}_t) = \NPvec_t && \label{eq:nrel-eb_zonal} \\
    & \NPvec_{t} = \EXvec_t \bm{e} &&     \label{eq:nrel-ex}\\
        &\bm{e}^T \bm{I}_t = 0 && \label{eq:nrel-balance}
    \end{align}%
\end{subequations}%

where $t$ indicates the market clearing time steps (e.g., hour or 15 minutes) and $T$ is the number of modeled timesteps.
Objective function \eqref{eq:nrel-obj} minimizes system cost given by the cost of generation $\mathcal{C}(\bm{G}_t)$ and the cost of curtailing \gls{res} $\mathcal{P}(\bm{C}_t)$, where $\mathcal{C}(\cdot)$ is a generator cost function model, $\bm{G}_t$ is the vector of generator production levels, $\bm{C}_t$ is the vector of \gls{res} curtailment, and $\mathcal{P}(\cdot)$ models the penalty for curtailment, e.g., by assigning a scalar penalty factor to the sum of curtailments.
Constraint \eqref{eq:nrel-capacity_g} enforces limits $\overline{\bm{g}}$ on generator outputs $\bm{G}_t$ and constraint \eqref{eq:nrel-capacity_c} limits curtailment $\bm{C}_t$ to the available (forecast) RES injection $\bm{r}_t$.
All inequalities on vectors are understood element-wise.
Nodal energy balance \eqref{eq:nrel-eb_nodal} defines nodal power injections $\bm{I}_t$ in terms of nodal load and generation.
For ease of notation and without loss of generality, we assume that each node hosts exactly one of each generator, load, and RES. This allows to model any node by adjusting $\overline{\bm{g}}$, $\bm{d_t}$, and $\bm{r}_t$ accordingly, e.g., setting $\overline{g_i}=0$ if node $i$ does not host a generator.
Similarly, the zonal energy balance defines the zonal net position $\NPvec_t$ as the difference between zonal load and generation by mapping resources and loads into each zone via map $\bm{m}\in\{0,1\}^{Z\times N}$.
Eq.~\eqref{eq:nrel-ex} defines auxiliary matrix $\EXvec_{t} \in\mathbb{R}^{Z\times Z} \ge \bm{0}$, which captures bilateral exchange between zones such that the element in the $k$-th row and $k'$-th column of $\EXvec_{t}$ defines the total flow of power from zone $k$ to zone $k'$. Vector $\bm{e}$ is the vector of ones in the appropriate dimensions.

Eq.~\eqref{eq:nrel-balance} enforces system balance. 
Note that all decision variable of the model are written in capital letters, while are parameters are given as lower case symbols. 

\subsubsection{Power transport limits}
Nodal power injections or zonal net positions of \gls{ed} \eqref{eqs:general_economic_dispspatch} can be subject to limitations given by the transmission system capacity and chosen power flow model.
Hence, FBMC-based market clearing can be modeled by using \gls{ed} \eqref{eqs:general_economic_dispspatch} and additionally enforcing
\begin{align}
    \NPvec_t \in \set{F}_t^{\text{zonal}} &\coloneqq \{ \bm{x} : \zPTDF_t \bm{x} \leq \RAM_t\} &&\forall t=1,...,T ,\label{eq:nrel-F_zonal}
\end{align}
where $\set{F}_t^\text{zonal}$ is the flow based domain as derived in Section~\ref{sec:nrel-fb_concept} above. 
Note that the zonal PTDF may be different for each time step, indicated by index $t$.
Alternatively, nodal market clearing, i.e., an ED that is constrained by all network transmission lines, can be modeled by constraining \eqref{eqs:general_economic_dispspatch} with 
\begin{align}
    \bm{I}_t \in \set{F}^\text{nodal} &\coloneqq \{ \bm{x} : {-\overline{\bm{f}} \le} \PTDF \bm{x} \leq \overline{\bm{f}}\} && \forall t=1,...,T. \label{eq:nrel-F_nodal}
\end{align}
Nodal market clearing limits the cross-zonal exchange only implicitly by taking into account the transmission capacity of the whole network. 
On the other hand, we can constrain cross-zonal exchange $\EX_{t,k,k`}$ directly using static bilateral \glspl{ntc}:
\begin{align}
\EXvec_t \in \set{F}_t^{\text{ntc}} \coloneqq &\{ x_{k,k'} : 0 \leq x_{k,k'} \leq \ntc_{k,k'}, \label{eq:nrel-F_ntc} \\   
& \qquad \forall k\neq k' = 1,...,E \} \hspace{1cm} \forall t=1,...,T, \nonumber
\end{align}
where $\ntc_{k,k'}$ is an externally defined parameter that limits the power exchange from zone $k$ to zone $k'$. Note that it is possible that $\ntc_{k,k'}\neq\ntc_{k',k}$ and that the approach in \eqref{eq:nrel-F_ntc} does not include a physical power flow model. 

\begin{table}
    \centering
    \caption[Model configuration for FBMC, nodal and NTC market clearing]{Model configuration for \gls{fbmc}, Nodal and \gls{ntc} market clearing.}
  \label{tab:nrel-model_composition}%
    \begin{tabular}{p{0.22\linewidth}p{0.2\linewidth}p{0.2\linewidth}p{0.18\linewidth}}
    \toprule
          & \textbf{FBMC}  & \multicolumn{1}{l}{\textbf{NTC}} & \textbf{Nodal} \\
    \midrule
    $D$--2:\newline Basecase & 
    \eqref{eq:nrel-obj} s.t.\newline \eqref{eq:nrel-capacity_g}--\eqref{eq:nrel-balance}, \eqref{eq:nrel-F_nodal}  &   --    & -- \\
    \midrule
    $D$--1:\newline Market Clearing & 
    \eqref{eq:nrel-obj} s.t.\newline \eqref{eq:nrel-capacity_g}--\eqref{eq:nrel-balance}, \eqref{eq:nrel-F_zonal} 
    &  \eqref{eq:nrel-obj} s.t.\newline \eqref{eq:nrel-capacity_g}--\eqref{eq:nrel-balance}, \eqref{eq:nrel-F_ntc}
    & \multirow[c]{2}{=}[-1.5em]{\eqref{eq:nrel-obj} s.t.\newline \eqref{eq:nrel-capacity_g}--\eqref{eq:nrel-balance}, \eqref{eq:nrel-F_nodal}} \\
    \cmidrule{1-3} 
    $D$--0:\newline Congestion\newline Management & 
    \multicolumn{2}{p{0.44\linewidth}}{\hspace{1.3cm}\eqref{eq:nrel-obj}+\eqref{eq:redisp_cost} s.t. \newline \hspace*{1.3cm}\eqref{eq:nrel-capacity_g}--\eqref{eq:nrel-balance}, \eqref{eq:nrel-F_nodal},\newline
    \hspace*{1.3cm} \eqref{eq:redisp_def}, \eqref{eq:redisp_curtlimit}}
          & \\
    \bottomrule
    \end{tabular}%
\end{table}%

\subsubsection{FBMC process}
The final FBMC process is computed as a three step sequence. Each step can be modeled through \eqref{eqs:general_economic_dispspatch} in combination with either \eqref{eq:nrel-F_zonal}, \eqref{eq:nrel-F_nodal} or \eqref{eq:nrel-F_ntc}.
Column ``FBMC'' of Table~\ref{tab:nrel-model_composition} itemizes the required modifications of \eqref{eqs:general_economic_dispspatch}.

Notably, the basecase is a nodal market clearing, following the intuition that the basecase should resemble $D$--0 as well as possible. 
The day-ahead market is cleared zonally with flow-based parameters $\zPTDF_t$ and $\RAM$ derived from the basecase results as per \eqref{eq:nrel-fb_parameters}.
$D$--0 congestion management, again, relies on nodal network representation \eqref{eq:nrel-F_nodal} and requires additional constraints that impose cost for deviating from the market clearing results:
\begin{subequations}\label{eqs:redispatch}
    \begin{align}
        \mathcal{C}(\bm{G}^{\text{red}}_t) &= (\bm{c}^{\text{red}})^T \bm{G}^{\text{red}}_t \label{eq:redisp_cost}\\
        \bm{G}_t - \bm{g}^{da}_t &= \bm{G}^{red}_t && \label{eq:redisp_def}\\
        \bm{C}_t &\geq \max\{0, \bm{r}_t - (\bm{r}_t^{da} - \bm{c}_t^{da})\} && \label{eq:redisp_curtlimit}
    \end{align}
\end{subequations}
where $\bm{g}_t^{da}$ and $\bm{c}_t^{da}$ are the decisions on $\bm{G}_t$ and $\bm{C}_t$ from the previous market clearing stage.
Vector $\bm{r}_t^{da}$ collects the day-ahead estimate of RES injections.
Note that we do not model an intraday market stage. 
Thus, all required re-dispatch actions are driven by the need to harmonize the zonal market outcomes with the actual grid situation in a nodal resolution.

\begin{remark}\label{rem:nodal_resolution}
We highlight that we model the D-2 stage as an economic dispatch with nodal resolution. We consider this the ``best estimate of the state of the system'' as required by the FBMC documentation \cite{50hertz_documentation_2020}. Further, this approach is most likely to resemble the actual TSO decision making process, which can rely on comprehensive historic data sets on the grid state.
Similar studies \cite{schonheit_minimum_2021} choose a basecase computation that internalizes \gls{fbmc} parameters in advance, e.g., by enforcing a given \gls{minram} requirement, force a zero balance state \cite{wyrwoll_determination_2019} or utilize zonal market clearing that does not consider network constraints \cite{matthes_impact_2019, finck_impact_2018}. This approach may bias the final \gls{fbmc} result, as shown in \cite{byers_modeling_2020}. 
\end{remark}

\begin{remark}
Note that Eq.~\eqref{eq:redisp_curtlimit} ensures that the RES injection limit defined at the DA stage is maintained in real time. 
While limit may be relaxed in real time at the discretion of the \glspl{tso} or by other (e.g., intraday) market stages, constraint \eqref{eq:redisp_curtlimit} ensures that the focus of the \gls{fbmc} re-dispatch stage is to adapt the day-ahead market outcome to the real-time grid situation with minimal re-dispatch, and not to find a new economically optimal market result. 
\end{remark}

\subsubsection{Reference formulations}
For reference, zonal market clearing using static bilateral \glspl{ntc} and a nodal market clearing are modeled and their resulting formulations are itemized in Table~\ref{tab:nrel-model_composition} as well.
The \gls{ntc} market clearing is modeled in two steps, because it does not require a basecase computation. 
The necessary congestion management step is the same as for \gls{fbmc}. 
The nodal market is a one-shot optimization of \eqref{eqs:general_economic_dispspatch} subject to nodal power flow constraints \eqref{eq:nrel-F_nodal}. Notably, the nodal market does not require a congestion management stage, because generation and network are co-optimized.

\subsection{Probabilistic FRMs via Chance Constraints}
\label{ssec:nrel-frm_cc}

The formulations of the previous section model \gls{fbmc} under perfect foresight of load and \gls{res} injections. 
The intention to provide efficient commercial exchange capacities to the market in combination with high shares of intermittent renewable generation poses the question of operability and how the forecasting characteristics of the basecase can be used to robustify results against \gls{res} uncertainty. 
As outlined in Section~\ref{sec:nrel-fb_concept} above, the \gls{fbmc} concept recognizes the existence of forecast uncertainties in the basecase by introducing \glspl{frm}, which are tuned based on historical data and \gls{tso}-defined \textit{risk levels} \cite[Fig.~4-2]{50hertz_documentation_2020}.
Specifically, \glspl{tso} use historical data to estimate the $(1-\epsilon)$-percentile of the absolute deviation between forecasted basecase flows $f^{bc}$, corrected by changes in the market schedule, and the realized real-time flows. By setting \glspl{frm} to at least the value of this $(1-\epsilon)$-percentile, \glspl{tso} ensure that lines are not overloaded due \gls{res} or load forecast errors with an empirical probability of $(1-\epsilon)$.
Thus, $\epsilon$ defines the risk level and is usually chosen small (e.g., $\epsilon = 5\%$).

\subsubsection{Motivation}

To date, studies on \gls{fbmc} have not considered \glspl{frm} in terms of an uncertainty model and risk-threshold, but rather employ fixed security margins that are applied uniformly to all \glspl{cne}, see e.g., \cite{schonheit_minimum_2021,schonheit_fundamental_2021,wyrwoll_determination_2019}.
While such simplifications may be motivated by a lack of historical data to simulate the \gls{frm} computation process prescribed by regulation, ignoring the specific impact of real-time control actions caused by intermittent renewable injections may obstruct a clear assessment of the effectiveness of \gls{fbmc} in \gls{res}-dominant systems.
As an alternative, we propose to model risk-aware \glspl{frm} that explicitly internalize \gls{res} uncertainty and the impact of real-time generator control actions on each \gls{cne}.
To this end, instead of creating an empirical uncertainty model of the flow forecast error on each \gls{cne}, we rely on a parameterized \gls{res} forecast error distribution and control participation factors. This approach leverages results from chance-constrained optimal power flow as proposed by \cite{bienstock_chanceconstrained_2014}.
\subsubsection{Formulation}

We model the uncertain injection from \gls{res} generators as $\bm{r}_t(\omega) = \bm{r}_t + \bm{\omega}_t$, where $\bm{\omega}_t$ is a zero-mean random vector that captures the forecast error of expected renewable generation $\bm{r}_t$.
We assume that the distribution of $\bm{\omega}_t$ can be modeled as a normal distribution such that $\bm{\omega}_t\sim \mathcal{N}(\bm{0},\bm{\Sigma}_t)$, where $\bm{\Sigma}_t$ denotes the covariance matrix of $\bm{\omega}_t$. 
Empirical results in \cite{dvorkin_uncertainty_2016} have shown that RES \textit{forecast errors} can indeed be well modeled using suitably parametrized normal distributions and we will follow this assumption in this paper.
However, as outlined in Remark~\ref{rem:dist_mod} below, the proposed approach can accommodate more relaxed distributional assumptions using the approaches discussed in \cite{roald_security_2015,dvorkin_chanceconstrained_2020}

Next, because the basecase and day-ahead markets are cleared based on forecast $\bm{r}_t$, error $\bm{\omega}_t$ will create a system imbalance. We assume that 
the control policy to restore system balance can be anticipated using the following model.
Similar to how \glspl{gsk} capture the estimated distribution of $\Delta\!\NPvec_t$ among all generators, we introduce vector of balancing participation factors $\bm{\alpha}_t$ that define the balancing control effort of generators as a response to imbalance $\bm{\omega}_t$ as:
\begin{align}
    \bm{G}_t(\bm{\omega}_t) = \bm{G}_t - \bm{\alpha}_t (\bm{e}^T\bm{\omega}_t).
    \label{eq:balancing_policy}
\end{align}
Since $\bm{\omega}_t$ and, thus, $\bm{G}_t(\bm{\omega}_t)$ are random variables, we first formulate the zonal \gls{ed} as a probabilistic problem
\allowdisplaybreaks
\begin{subequations}\label{eqs:expected_economic_dispspatch}%
    \begin{align}
        \min \quad & \mathbb{E}[\sum_{t=1}^T \mathcal{C}(\bm{G}_t(\bm{\omega}_t))] \label{eq:nrel-exp_cc_obj} \\ 
        \text{s.t.} \quad & \forall t=1,...,T: \nonumber  \\
        &\mathbb{P}[0 \leq G_{t,i}(\bm{\omega}_t) \leq \overline{g}_i] \geq 1 - \epsilon &&  \hspace{-2cm} \forall i=1,...,N \label{eq:nrel-cc_gencons}\\
        &\mathbb{P}[\zPTDF_{t,j} \!\cdot\! \NPvec_t(\bm{\omega}_t) \leq \overline{f}_{j} - f_{t,j}^{ref}] \geq 1 - \epsilon \hspace{-4cm}&& \nonumber \\
        & &&\hspace{-2cm}\forall j=1,...,E \label{eq:nrel-cc_flowcons} \\
        &\bm{m}(\bm{G}_t(\bm{\omega}_t) + \bm{r}_t(\bm{\omega}_t) - \bm{C}_t - \bm{d}_t) = \NPvec_t(\bm{\omega}_t) && \nonumber \\
        & &&\hspace{-2cm} \forall\bm{\omega}_t\!\in\! \Omega_t, \label{eq:nrel-cc_balance} \\
        &\bm{e}^T \NPvec_t(\bm{\omega}_t)  = 0 && \hspace{-2cm}\forall\bm{\omega}_t\!\in\! \Omega_t \label{eq:nrel-cc_system_balance}
    \end{align}%
\end{subequations}%
\allowdisplaybreaks[0]%

where $E$ is the number of \glspl{cnec} and $\Omega_t$ denotes the set of all possible outcomes of $\bm{\omega}_t$.
Objective \eqref{eq:nrel-exp_cc_obj} minimizes the expected system cost. 
Constraints \eqref{eq:nrel-cc_gencons} and \eqref{eq:nrel-cc_flowcons}, ensure that the probability that a generator can fulfill its required response $\bm{\alpha}_t (\bm{e}^T\bm{\omega}_t)$ or a \gls{cnec} $j$ is not overloaded is at least $(1-\epsilon)$. 
These so called \textit{chance constraints} resemble the value-at-risk, a risk metric commonly used in the finance industry \cite{bienstock_chanceconstrained_2014}.
Lastly, Eqs.~\eqref{eq:nrel-cc_balance} and \eqref{eq:nrel-cc_system_balance} ensure that the system is balanced for all possible outcomes of $\bm{\omega}_t \in \bm{\Omega}_t$. 

Problem \eqref{eqs:expected_economic_dispspatch} can not be solved directly, but allows a computationally tractable deterministic reformulation. 
First, recall that $\bm{\omega}_t$ is zero mean, i.e., $\mathbb{E}[\bm{\omega}_t]=0$. 
For a linear cost function model $\mathcal{C}(\bm{G}_t(\bm{\omega}_t)) = \bm{R}^T \bm{G}_t(\bm{\omega}_t)$, where $\bm{R}^T$ is a vector of cost factors, we therefore get $\mathbb{E}[\sum_{t=1}^T \mathcal{C}(\bm{G}_t(\bm{\omega}_t))] = \bm{R}^T \bm{G}_t$ as per \eqref{eq:balancing_policy}.
We will use this linear cost model for the remainder of this paper and refer to, e.g., \cite{mieth_risk_2020} for analogous derivations for quadratic generator cost models. 

Next, chance-constraints \eqref{eq:nrel-cc_gencons} and \eqref{eq:nrel-cc_flowcons} can be reformulated by recalling that for any normally distributed random variable $x\sim \mathcal{N}(\mathbb{E}[x],\sigma(x))$ it holds that:
\begin{equation}
    \mathbb{P}[x\le \overline{x}] \ge (1-\epsilon) \quad \Leftrightarrow \quad \mathbb{E}[x] + \Phi^{-1}(1-\epsilon) \sigma(x) \le \overline{x},
    \label{eq:general_cc_reform}
\end{equation}
where $\Phi$ is the cumulative distribution function of the standard normal distribution and $\sigma(x)$ is the standard deviation of $x$.
Eq.~\eqref{eq:general_cc_reform} indicates that we require expected values and standard deviations to reformulate \eqref{eq:nrel-cc_gencons} and \eqref{eq:nrel-cc_flowcons}. Thus, we compute:  
\allowdisplaybreaks
\begin{align}
    &\mathbb{E}[G_{t,g}(\bm{\omega}_t)] = \mathbb{E}[G_{t,g} - \alpha_{t,g}(e^T\bm{\omega}_t)] = G_{t,g}\\
    \begin{split}
        &\sigma(G_{t,g}(\bm{\omega}_t)) = \sqrt{\Var[\alpha_{t,g}(e^T\bm{\omega}_t)]} \\
        &\phantom{\sigma(G_{t,g}(\bm{\omega}_t))}= \sqrt{\alpha_{t,g}^2 (e^T \Sigma_t e)} =  \alpha_{t,g} s_t 
    \end{split}\\ \nonumber \\
    \begin{split}
        &\mathbb{E}[\zPTDF_{t,j} \cdot \NPvec_t(\bm{\omega}_t)] \\
            &\quad= \mathbb{E}[\zPTDF_{t,j}(m (\bm{G}_t(\bm{\omega}_t)) + \bm{r}_t(\bm{\omega}_t)) - \bm{d}_t)] \\
            &\quad= \zPTDF_{t,j}(m(\bm{G}_t + \bm{r}_t - \bm{d}_t)
    \end{split}\\ \nonumber \\
    \begin{split}
        &\sigma[\zPTDF_{t,j} \cdot \NPvec_t(\bm{\omega}_t))] \\
            &= \sqrt{\Var[\zPTDF_{t,j}(m(\bm{G}_t(\bm{\omega}_t) + \bm{r}_t(\bm{\omega}_t) - \bm{d}_t)]} \\
            &= \sqrt{(\zPTDF_{t,j} \bm{\alpha}_t \bm{e}^T) \bm{\Sigma}_t (\PTDF_{t,j} \bm{\alpha}_t \bm{e}^T)^T} \\
            &= \lVert (\zPTDF \bm{\alpha}_t \bm{e}^T) \bm{\Sigma}^{1/2}_t \rVert_2,
    \end{split}%
\end{align}%
\allowdisplaybreaks[0]%
where we define $s^2_t = \bm{e}^T \bm{\Sigma}_t \bm{e}$ and $\lVert \cdot \rVert_2$ denotes the 2-norm. 
Lastly, \eqref{eq:nrel-cc_balance} holds for all $\bm{\omega}_t$, if the system is balanced in expectation and the sum of all control actions is exactly equal to the system imbalance, i.e.,:
\begin{align}
    \bm{e}^T \bm{\alpha}_t (\bm{e}^T\bm{\omega}) = \bm{e}^T \bm{\omega}_t \quad \Leftrightarrow \quad \bm{e}^T \bm{\alpha}_t = 1. 
\end{align}
Thus, the deterministic reformulation of \eqref{eqs:expected_economic_dispspatch} is given as:
\allowdisplaybreaks
\begin{subequations}\label{eqs:cc_economic_dispspatch}%
    \begin{align}
    \min \quad &\sum_{t=1}^T \bm{R}^T\bm{G}_t \quad s.t. && \label{eq:nrel-cc_obj}\\
    \text{s.t.} \quad & \forall t=1,...,T \nonumber  \\
    &\bm{G}_t + z_{\epsilon} s_t \bm{\alpha}_t \leq \overline{\bm{g}}  \\
    & -\bm{G}_t + z_{\epsilon} s_t \bm{\alpha}_t \geq \bm{0} \\
    &\zPTDF_{t,j} \NPvec_t \leq \overline{f}_{j} - f_{t,j}^{ref} - z_{\epsilon} T_{t,j}  &&\forall j=1,...,E \label{eq:nrel-cc_domain}\\ 
    &\lVert (\zPTDF_{t,j} \bm{\alpha}_t \bm{e}^T) \bm{\Sigma}^{1/2} \rVert_2 \leq T_{t,j} && \forall j=1,...,E \label{eq:nrel-cc_soc}\\
    &\bm{m}(\bm{G}_t + \bm{r}_t - \bm{d}_t) = \NPvec_t \\
    & \bm{e}^T \bm{\alpha}_t = 1 \\
    & \bm{e}^T \NPvec_t  = 0 
    \end{align}%
\end{subequations}%
\allowdisplaybreaks[0]%
where we use $z_{\epsilon} = \Phi^{-1}(1-\epsilon)$ for a more concise notation and introduce auxiliary variable $T_{t,j}$ to denote the standard deviation of the flow across \gls{cnec}~$j$.
Constraint \eqref{eq:nrel-cc_soc} is a second-order conic constraint.
As a result, \eqref{eqs:cc_economic_dispspatch} becomes a convex second-order conic program, which can be solved directly and efficiently by many off-the-shelf solvers.%
\footnote{A useful list of solvers and the problem types they support can be found here: \url{https://jump.dev/JuMP.jl/stable/installation/\#Install-a-solver}.}
To facilitate the solution of very large problem instances, additional cutting-plane algorithms as in \cite{bienstock_chanceconstrained_2014} can be applied.
The variable $\bm{\alpha}_t$ can be chosen as parameter, similarly to the \gls{gsk}, or optimized as a decision variable in \eqref{eqs:cc_economic_dispspatch}. 
In this paper we use the latter approach, thus allowing for an optimized generator response. 

\begin{figure}
    \centering
    \includegraphics[width=0.9\linewidth]{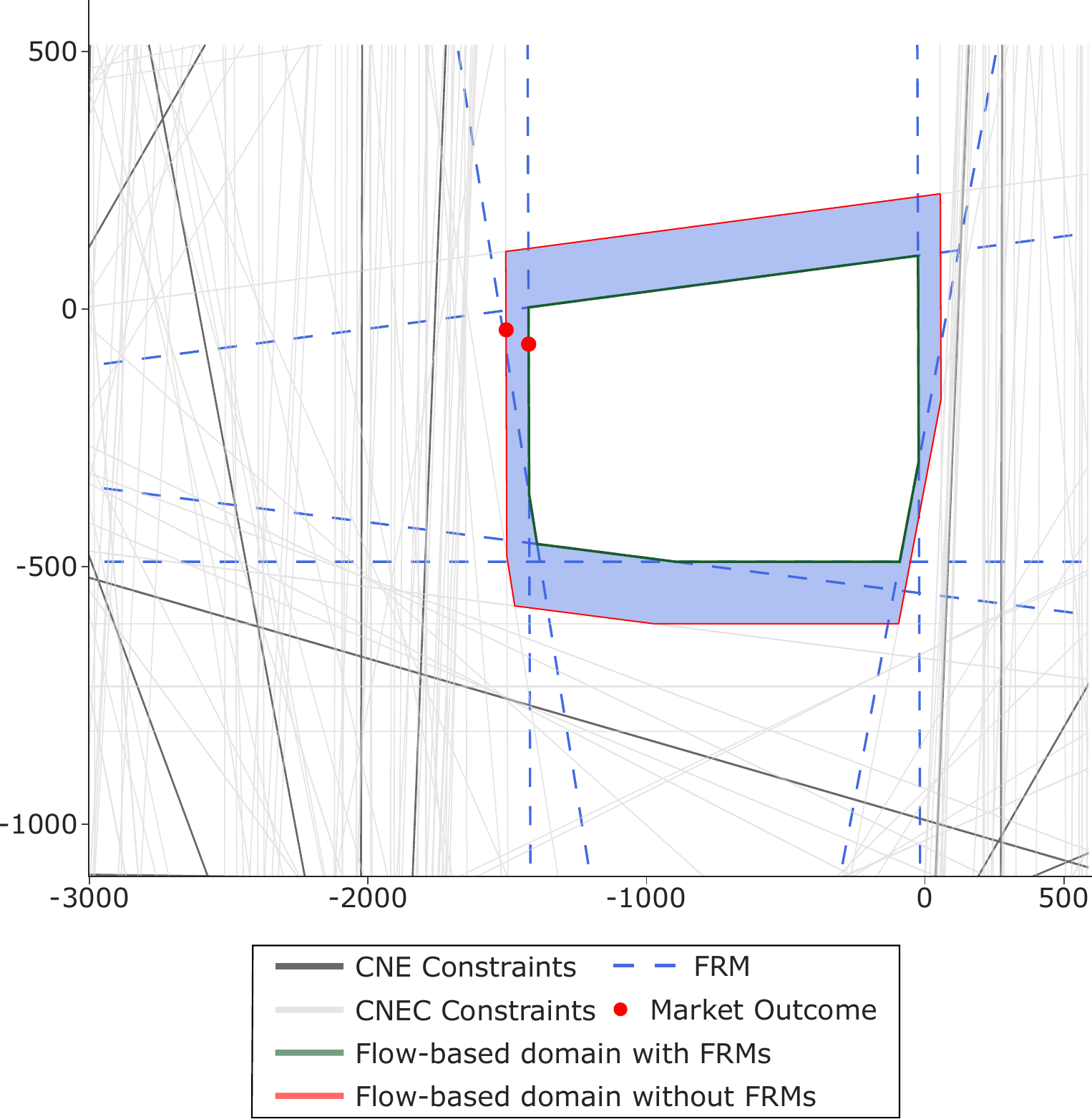}
    \caption[Flow-based domain with including FRMs]{Flow-based domain with \glspl{frm} for exchange \textit{Zone 1} - \textit{Zone 2} (x-axis) and \textit{Zone 2} - \textit{Zone 3} (y-axis).}
    \label{fig:nrel-domain_with_frm}
\end{figure}

\begin{remark}\label{rem:dist_mod}
Program \eqref{eqs:cc_economic_dispspatch} uses a parametric uncertainty model, i.e., it can be solved as a deterministic one-shot program. 
This uncertainty model is estimated beforehand by the \gls{tso} by computing the parameters of suitable probability distribution functions, e.g., as described in \cite{dvorkin_uncertainty_2016}.
In our case, we assume a zero-mean normal distribution with covariance matrix $\bm{\Sigma}_t$.
This assumption can be relaxed by re-defining $z_\epsilon$ in \eqref{eqs:cc_economic_dispspatch}. For example, for modeling a zero-mean Student's $t$ distribution with $\nu$ degrees of freedom and scale parameter $\sigma_T = (\nu-2)/\nu$ define $z_\epsilon \coloneqq t_{\nu,\sigma_T}^{-1}(1-\epsilon)$ with $t_{\nu,\sigma_T}$ being the Student's cumulative distribution function. For a distributionally robust approach that only uses moment information (mean and variance), define $z_\epsilon = \sqrt{(1-\epsilon)/\epsilon}$ \cite{roald_security_2015}.
\end{remark}

By using \eqref{eqs:cc_economic_dispspatch} to clear the D-1 market stage, we can interpret term $z_{\epsilon} T_{t,j}$ in Eq.~\eqref{eq:nrel-cc_domain} as a line-specific endogenous \gls{frm} that reduces capacity for each \gls{cnec} based on the \gls{res} uncertainty model. 
Through the chance-constraints in \eqref{eq:nrel-cc_gencons} and \eqref{eq:nrel-cc_flowcons}, the proposed approach remains similar to current TSO practice of expressing uncertainty in terms of risk-levels. 
However, instead of enforcing fixed transmission line security margins derived from historical flow changes, it explicitly internalizes the current grid situation, dispatch decisions, and a model of the generator balancing policy expressed by $\bm{\alpha}_t$.
At the same time, it avoids common caveats of alternative approaches such as scenario-based or robust optimization. First, the chance-constrained optimal power flow formulation in \eqref{eqs:cc_economic_dispspatch} is a deterministic single-level problem. As a result, it is more computationally tractable than scenario-based approaches and is insensitive to scenario selection and generation. Second, by allowing the decision maker to control the desired risk level it avoids overly conservative, and hence costly, solutions created by robust approaches. We refer the interested reader to \cite{bienstock_chanceconstrained_2014,dvorkin_chanceconstrained_2020,mieth_risk_2020} for further discussion.
Finally, we highlight that, on the one hand, the proposed approach may be of interest to \glspl{tso} due to the similarity of the chance constraints to the current practice and the similarity of the generator balancing model to the existing \glspl{gsk}. On the other hand, the proposed approach provides a suitable method for independent FBMC modelers without access to detailed historical data to compute \glspl{frm} more accurately.

\subsubsection{Illustration}

Consider Fig.~\ref{fig:nrel-domain_with_frm} for an illustration of the impact of the \glspl{frm} derived from chance constraints on the flow-based domain.
This figure shows the flow-based domain with and without \glspl{frm} as two polygons for a single time step and the exchange between 3 zones. The x-axis shows the net exchange from \textit{Zone 1} to \textit{Zone 2} and the y-axis shows the net exchange from \textit{Zone 2} to \textit{Zone 3}.
Details on the specific data used for the underlying computations is provided in Section~\ref{sec:nrel-case_study}.

Each line in Fig.~\ref{fig:nrel-domain_with_frm} corresponds to a \gls{cnec} (shown in light gray) or \gls{cne} (shown in dark gray). 
The combination of the \glspl{cne} and \glspl{cnec} that constrain the net exchanges between the three zones the most define the flow-based domain highlighted in red. 
The outer, larger red polygon captures the flow-based domain based on forecast $\bm{r}_t$ and without any security margin.
The blue dashed lines shown the most constraining \glspl{cne} and \glspl{cnec} \textit{with} the endogenous \glspl{frm} and the inner green polygon shows the resulting, smaller, flow-based domain.
On the one hand, it is clear that security margins on CNEs and CNECs reduce the flow-based domain and the resulting market outcome (shown as a red dot) will also move inward, leading to lower potential commercial exchange. 
On the other hand, Fig.~\ref{fig:nrel-domain_with_frm} highlights, that the necessary margin (i.e., the distance between the gray and blue lines of the most constraining \glspl{cne} and \glspl{cnec}) to achieve the desired $(1-\epsilon)$-security differs between the various \glspl{cne} and \glspl{cnec}.
This indicates that a fixed \gls{frm} proxy margin would over- or underestimate the \gls{ram} on some \glspl{cne} and \glspl{cnec}.
Note that the selection of \glspl{cne} and \glspl{cnec} only depends on the sensitivity of network elements towards interzonal flow (see Section~\ref{sssec:cne_cnec} above). This sensitivity is defined by $\PTDF$ and $\zPTDF$, which remain constant if network topology and GSK strategy remains constant. 

\section{Case Study}\label{sec:nrel-case_study}

\subsection{Data Set}

\begin{figure}
  \centering
    \includegraphics[width=0.8\linewidth]{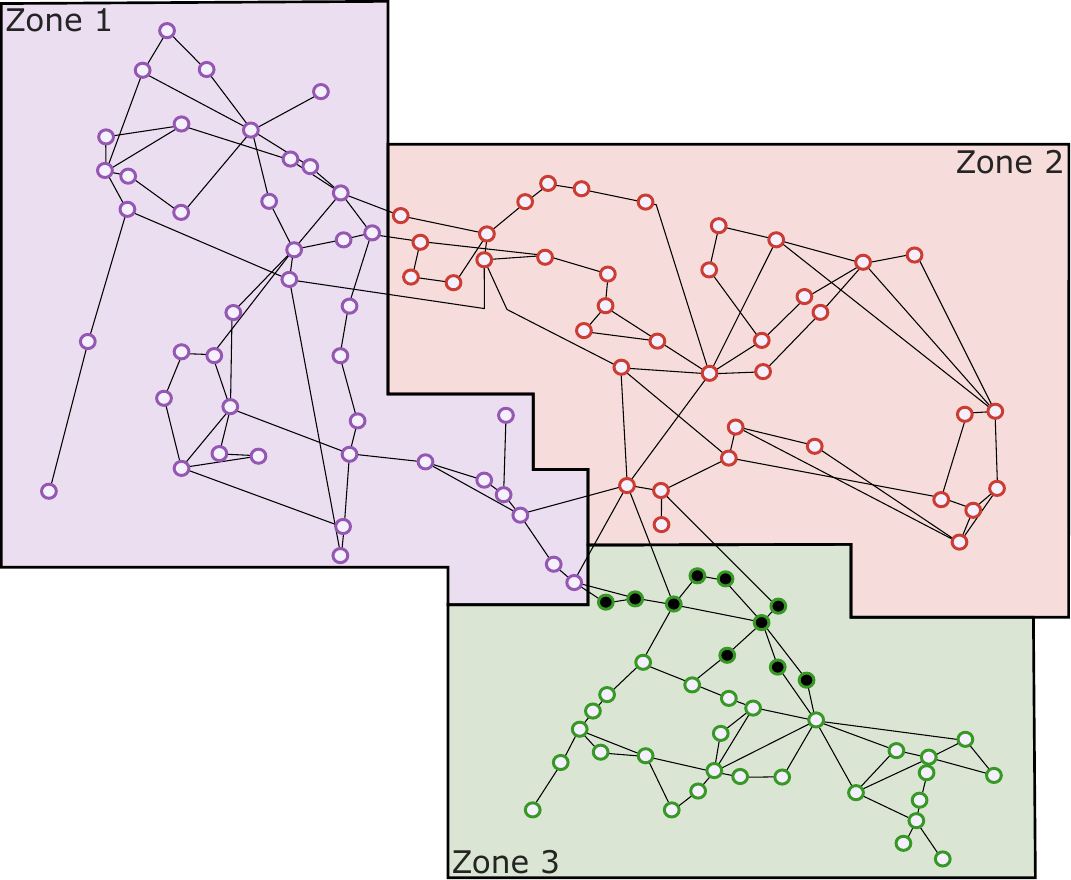}
  \caption{Topology of the IEEE 118 bus system, zones indicated by color. Changes in topology from the original data are indicated with filled nodes. }
  \label{fig:nrel-map}
\end{figure}

The numerical experiments use the IEEE 118 bus test system as published in \cite{pena_extended_2017}, which augments this data set by additional generation technologies, a three-zone configuration, and hourly load and \gls{res} injection timeseries for a full year. In \cite{pena_extended_2017} the authors kept the original topology from the original IEEE 118 bus network, but line capacities are generously (4x-5x) scaled with installed capacity.
We chose this well-known and -documented data set for our experiments to facilitate reproducibility and interoperability of our results.

\begin{figure}
  \centering
    \includegraphics[width=0.95\linewidth]{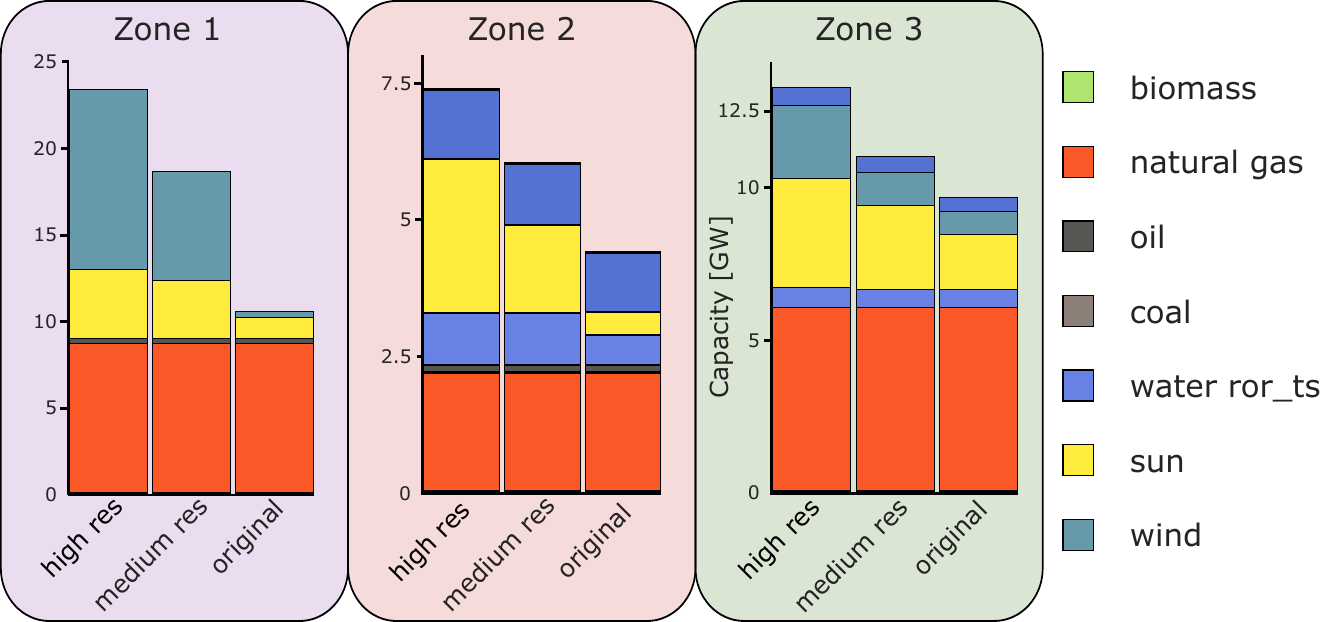}
  \caption{Installed capacities for the three scenarios \textit{original}, \textit{medium res} and \textit{high res}.}
  \label{fig:nrel-installed_capacities}
\end{figure}

The original data in \cite{pena_extended_2017} is complemented in this study with two scenarios that further increase the share of \gls{res} generation in the total generation over the model horizon from 26\% in the original data to 50\% (\textit{medium res}) and 70\% (\textit{high res}). The resulting installed capacities are itemized in Fig.~\ref{fig:nrel-installed_capacities} for the three scenarios \textit{original}, \textit{medium res} and \textit{high res}.
To better reflect the scarcity of transmission capacity, all line capacities are scaled down by 30\%.
Additionally, the zonal configuration was slightly adjusted so all zones have shared borders. 
This is indicated in Fig.~\ref{fig:nrel-map} by nodes that are filled solid black, which where allocated to ``Zone 2'' (red) and are now allocated to ``Zone 3'' (green). 

We analyze the effectiveness of \gls{fbmc} by comparing system cost, which are composed of generation cost at the market clearing stage and additional congestion management (redispatch) in $D$--0. See also \eqref{eqs:redispatch}. 
We set the cost for redispatch to 30\$ per MWh and curtailment cost to 5\$ per MWh.  
Note that congestion management does not just revert a zonal solution to an optimal nodal solution, but tries to achieve a network-feasible solution with minimal deviations from the zonal solution. 

\subsection{Zonal benchmarks}

\begin{figure}
  \centering
    \includegraphics[width=0.99\linewidth]{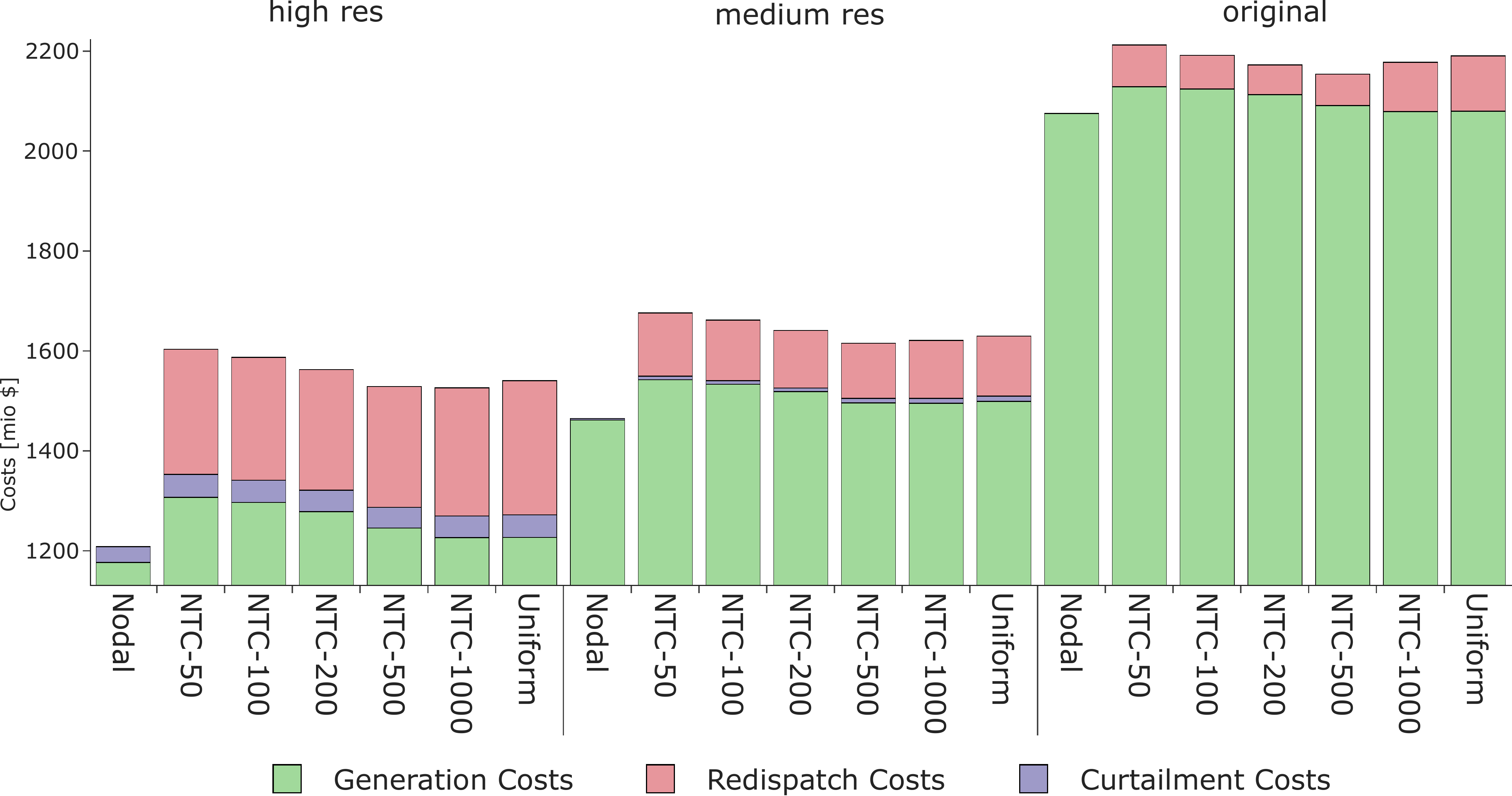}
  \caption{Cost composition of zonal \glspl{ntc} market clearing and the nodal reference. Note that Generation Cost capture the total cost after redispatch.}
  \label{fig:nrel-ntc_nodal_reference}
\end{figure}

Fig.~\ref{fig:nrel-ntc_nodal_reference} shows the benchmark reference results for the three scenarios for nodal market clearing and zonal market clearing subject to \glspl{ntc}. See also Table~\ref{tab:nrel-model_composition}. 
System cost are evaluated after congestion management and include cost for curtailment and redispatch. The Nodal solution on the left represents the economic optimum that does not require redispatch. 
Zonal market clearing is shown with increasing \glspl{ntc} from left to right. 
The ``Uniform pricing`` case describes market clearing without any commercial exchange constraints.

The results show the expected pattern: 
\begin{itemize}
    \item Generation cost decrease with higher shares of cheap generation from \gls{res}.
    \item The nodal dispatch is the least cost solution, with no redispatch required. Additionally, due to cost for congestion management, zonal solutions often show increased generation cost. 
    \item For zonal market clearing, more exchange capacities lead to lower generation cost. However, cost for congestion management can outweigh these savings.
    \item Zonal D-1 market clearing leads to higher overall generation cost compared to the Nodal solution, because minimizing generation schedule changes in D-0 are prioritized over minimizing generation cost.
\end{itemize}
Notably, \glspl{ntc} of 500~MW lead to the lowest cost dispatch for zonal market clearing. In the following, this result will be used as a benchmark for the \gls{fbmc} results. 

\subsection{Flow-based parametrization}
\label{ssec:nrel-fb_parameters}

As described in Section~\ref{sec:nrel-formal_fbmc}, the flow-based parameters are used to solve the day-ahead stage in FBMC. 
They are composed of the zonal \gls{ptdf} and \gls{ram} values and calculated from basecase generation schedules $\bm{g}_t^{bc}$, power flows $\bm{f}_t^{bc}$ and net positions $\NPvec_t^{bc}$.
As discussed in Remark~\ref{rem:nodal_resolution} above, we use nodal pricing with full network representation (case Nodal) to stay as close as possible to the reference given by the documented process in~\cite{50hertz_documentation_2020}.

The zonal \gls{ptdf} is composed of all cross-border lines and internal lines with a zone-to-zone \gls{ptdf} value larger than 5\%. Contingencies are included based on a 20\% line-to-line sensitivity in case of an outage using so called load-outage distribution factors as per \cite{jiachun_direct_2009}, i.e. lines are considered contingencies that distribute 20\% of line loading to the \gls{cne}. 

To focus the analysis on the permissiveness of \gls{fbmc} in the context of RES injections, all scenarios use a so called \textit{Pro-Rata} approach to calculate \glspl{gsk}, i.e., changes in net position are distributed based on the online dispatchable generation capacity at each time step. This approach represents a suitable baseline as it is employed in practice \cite{50hertz_documentation_2020} as well as described in academic literature \cite{dierstein_impact_2017}.

The \gls{minram} parameter is chosen to reflect a restrictive and permissive parametrization of flow-based parameters. First, consistent with practice \cite{amprion_amprion_2019} and the \gls{minram} value for coordinated capacity allocation within the flow-based region\cite[Article 17]{acer_2019}, a 20\% \gls{minram} is
employed. This also ensures feasibility of \eqref{eq:nrel-fb_parameters_3} as the feasible region \eqref{eq:nrel-F_zonal} has cases where a suboptimal \gls{gsk} leads to negative \gls{ram} values. The resulting \gls{fbmc} configuration is denoted as \gls{fbmc} in the following paragraphs. 
Second, to illustrate the impact of less restrictive flow-based parameterization we analyse a case that, consistent with regulation, enforces a \gls{minram} of 70\%\cite{europeancommission_commission_2019} and only considers cross-border lines as \glspl{cne}\cite[Article 5]{acer_2019}. We denote this case as \textit{FBMC$^+$}.

All scenarios and market configurations where solved using the open \glsreset{pomato}\gls{pomato} \cite{weinhold_power_2020} written in Python and Julia. All computations have been performed on standard PC hardware with a Ryzen 7 processor and 32GB of memory using the Gurobi solver \cite{gurobioptimizationllc_gurobi_2018} for the deterministic and chance constrained formulation.
The computation times for a single day were around 0.5 seconds for the deterministic $D$--1 stage, around 19 seconds for the chance-constrained $D$--1 stage, and around 1.6 seconds for the $D$--0 stage.

Note that the computational complexity of the proposed approach depends on the number of considered zones and identified \glspl{cnec}, which constrain the market and generators that are considered for balancing control.
See also Section~\ref{ssec:scalability} for additional results for a larger system.

\subsection{Deterministic FBMC}

Table~\ref{tab:nrel-cost} shows the total system costs for the three scenarios. \textit{Nodal} represents the first-best reference and \textit{NTC-500} is included as the best zonal reference.

The additional cost for congestion management depend on the system's ability to accommodate the market result. Here the \textit{\gls{fbmc}} and \textit{\gls{ntc}} solutions are very close, with \textit{\gls{fbmc}} generally resulting in higher generation cost, but lower cost for congestion management. Similarly to the reference results from Section~\ref{sec:nrel-case_study}, the \gls{ntc} and \gls{fbmc} solutions illustrate the trade-off between capacity allocation and congestion management, where \gls{fbmc} is more restrictive and generally leads to less congestion management at higher cost in generation. 

\begin{table}
    \centering
    \caption[System cost including generation and congestion management.]{System cost including generation and congestion management (CM) for each scenario.}
    \label{tab:nrel-cost}%
    \begin{tabular}{clllll}
    \toprule
          &       & \multicolumn{1}{l}{FBMC} & \multicolumn{1}{l}{FBMC$^+$} & \multicolumn{1}{l}{NTC-500 } & \multicolumn{1}{l}{Nodal } \\
    \midrule
    \multirow{5}[4]{*}{original} & Generation & 2091.45 & 2083.82 & 2091.01 & 2075.16 \\
          & Curtailment & 0.03  & 0.08  & 0.01  & 0 \\
          & Redispatch & 48.76 & 89.28 & 62.89 & 0 \\
\cmidrule{2-6}          & total CM & 48.79 & 89.36 & 62.9  & 0 \\
          & total & 2140.24 & 2173.18 & 2153.92 & 2075.16 \\
    \midrule
    \multirow{5}[4]{*}{medium res} & Generation & 1498.85 & 1499.01 & 1496.28 & 1461.68 \\
          & Curtailment & 7.8   & 9.97  & 8.55  & 3.06 \\
          & Redispatch & 110   & 115.91 & 110.61 & 0 \\
\cmidrule{2-6}          & total CM & 117.8 & 125.88 & 119.17 & 3.06 \\
          & total & 1616.66 & 1624.89 & 1615.44 & 1464.74 \\
    \midrule
    \multirow{5}[4]{*}{high res} & Generation & 1256.1 & 1239.05 & 1245.6 & 1176.48 \\
          & Curtailment & 42.1  & 42.73 & 41.57 & 31.81 \\
          & Redispatch & 234.89 & 243.57 & 241.63 & 0 \\
\cmidrule{2-6}          & total CM & 276.99 & 286.29 & 283.2 & 31.81 \\
          & total & 1533.09 & 1525.35 & 1528.8 & 1208.29 \\
    \bottomrule
    \end{tabular}%
\end{table}%

The results from the more permissive \textit{FBMC$^+$} configuration, which enforces a 70\% \gls{minram} and only considers cross-border \glspl{cne}, further illustrates this point. 
Table~\ref{tab:nrel-cost} shows the cost decomposition and Table~\ref{tab:nrel-congestion_management} show the respective redispatch (column R), curtailment (column C) and combined congestion management (column C+R) volumes for each scenario and include the \textit{FBMC$^+$} configuration. 
Here, the \textit{FBMC$^+$} proves less restrictive than the \gls{ntc} and \gls{fbmc} configurations. With the original data, the relaxed flow-based parameters \textit{FBMC$^+$} lead to overall higher cost, due to increased congestion management. For higher shares of intermittent renewable generation, larger exchange capacities become more efficient and, while still with the higher congestion management volumes, provide the lowest cost for zonal market clearing.  

\begin{table}
    \centering
    \caption[Quantities of redispatch and curtailment in TWh.]{Quantities of redispatch (R) and curtailment (C) in TWh for each scenario.}
    \label{tab:nrel-congestion_management}%
    \resizebox{\linewidth}{!}{
    \begin{tabular}{l|lll|lll|lll}
    \toprule
          & \multicolumn{3}{c|}{original} & \multicolumn{3}{c|}{medium res} & \multicolumn{3}{c}{high res} \\
          & C     & R     & C+R & C     & R     & C+R & C     & R     & \multicolumn{1}{l}{C+R} \\
\cmidrule{1-10}    FBMC  & 0.01  & 1.63  & 1.64  & 1.56  & 3.67  & 5.23  & 8.42  & 7.83  & 16.25 \\
    FBMC$^+$ & 0.02  & 2.98  & 3     & 1.99  & 3.86  & 5.85  & 8.55  & 8.12  & 16.67 \\
    NTC-500 & 0     & 2.1   & 2.1   & 1.71  & 3.69  & 5.4   & 8.31  & 8.05  & 16.36 \\
    Nodal & 0     & 0     & 0     & 0.61  & 0     & 0.61  & 6.36  & 0     & 6.36 \\
    \bottomrule
    \end{tabular}%
    }
    \vspace{-1em}
\end{table}%

\subsection{Probabilistic FRMs}

The \textit{high res} scenario reaches a share of intermittent renewable generation of 60\%. The dispatch of such high RES injections proves challenging when accounting for deviations between the day-ahead market stage and real-time. In Section~\ref{ssec:nrel-frm_cc} we propose a chance-constrained formulation for \glspl{frm}, that reduce the flow-based domain based on an assumed distribution of forecast errors $\bm{\omega}_t$.
Following related publications \cite{dvorkin_chanceconstrained_2020,mieth_risk_2020} we set the system risk-level to $\epsilon=5\%$ and calculate $\bm{\Sigma}_t$ by assuming the standard deviation of each forecast error to be $10\%$ and without correlation among each other.
Note that if historical data is available $\bm{\Sigma}_t$ can also be estimated directly from this data.

The results in the previous section illustrate the trade-off between permissive capacity allocation and increased congestion management, which can be desirable depending of the associated cost. 
However, the cost for congestion management will change, and presumably increase, if the real-time availability is subject to forecast errors. For this numerical experiment we take a closer look at the \textit{FBMC$^+$} scenario, as it provides the largest trading capacities and is most sensitive to forecast errors. 

We consider the case \textit{FBMC$^+$ CC} that now includes risk-aware \glspl{frm} into the \textit{FBMC$^+$} case by solving \eqref{eqs:cc_economic_dispspatch} in the day-ahead market clearing stage.
As discussed in Section~\ref{ssec:nrel-frm_cc} above, this will reduce the transfer capacities that are allocated to the market.
To evaluate the impact of forecast errors that occur in real time on congestion management, the final FBMC stage is then run 20 times for both \textit{FBMC$^+$ CC} and \textit{FBMC$^+$} using outcomes of $\bm{\omega}_t$ drawn from the distribution described above.

The generator response $\bm{\alpha}_t$, that is an endogenous result from the chance constraint formulation, is used for both scenarios to calculate the generator response real-time deviations. The numeric results are obtained from 20 full-year runs, with hourly independent real-time deviations, leading to a total of 7,300 computed daily scenarios.  

Table~\ref{tab:nrel-cc_cm} shows that, indeed, the resulting expected cost for congestion management and the overall system cost are lower in the \textit{FBMC$^+$ CC} if \gls{res} uncertainty is explicitly considered.
Here, we see the same values as in Table~\ref{tab:nrel-cost} with no deviations at real-time $\omega = 0$ with \textit{FBMC$^+$ CC} resulting in higher cost for congestion management. With real-time deviations $\omega > 0$ cost for congestion management are overall lower in the \textit{FBMC$^+$ CC} due to reduced exchange margins in response of expected deviations and the set generator's response. 

The results in Table~\ref{tab:nrel-cc_cm} allow a \textit{regret analysis} (see, e.g., \cite{mieth2019online}), i.e., we can compute how much more \textit{FBMC$^+$ CC} reduces the cost of uncertainty 
 than \textit{FBMC$^+$} in relation to the theoretical system cost without uncertainty---marked $(i)$ in Table~\ref{tab:nrel-cc_cm}. 
Using the values in the cells labeled $(i)$--$(iii)$ in Table~\ref{tab:nrel-cc_cm}, the regret of \textit{FBMC$^+$} is $(ii)-(i)=3.54$, and the regret of \textit{FBMC$^+$ CC} is $(iii)-(i)=2.58$ indicating a 27\% improvement from the proposed method.

\begin{table}[t]
\setlength{\tabcolsep}{5pt}
  \centering
    \caption{Cost for redispatch (R) and curtailment (C) in relation to forecast error $\omega$.}\label{tab:nrel-cc_cm}%
    \begin{tabular}{llll|lll}
    \toprule
          & \multicolumn{3}{c|}{$\omega = 0$} & \multicolumn{3}{c}{$\omega > 0$} \\
          & \multicolumn{1}{l}{C} & \multicolumn{1}{l}{R} & \multicolumn{1}{l|}{C+R} & \multicolumn{1}{l}{C} & \multicolumn{1}{l}{R} & \multicolumn{1}{l}{C+R} \\
    \midrule
    FBMC$^+$ & 42.73 & 243.57 & 286.29\textsuperscript{$(i)$} & 41.42 & 248.41 & 289.83\textsuperscript{$(ii)$} \\
    FBMC$^+$ CC & 44.43 & 246.6 & 291.03 & 42.09 & 245.16 & 287.25\textsuperscript{$(iii)$} \\
    \bottomrule
    \end{tabular}%
\flushleft
$(i)$ Theoretical optimal system cost without uncertainty.\newline
$(ii)$ System cost with uncertainty without the proposed method.\newline
$(ii)$ System cost with uncertainty with the proposed method.
\end{table}%

\begin{figure}
  \centering
    \includegraphics[width=1\linewidth]{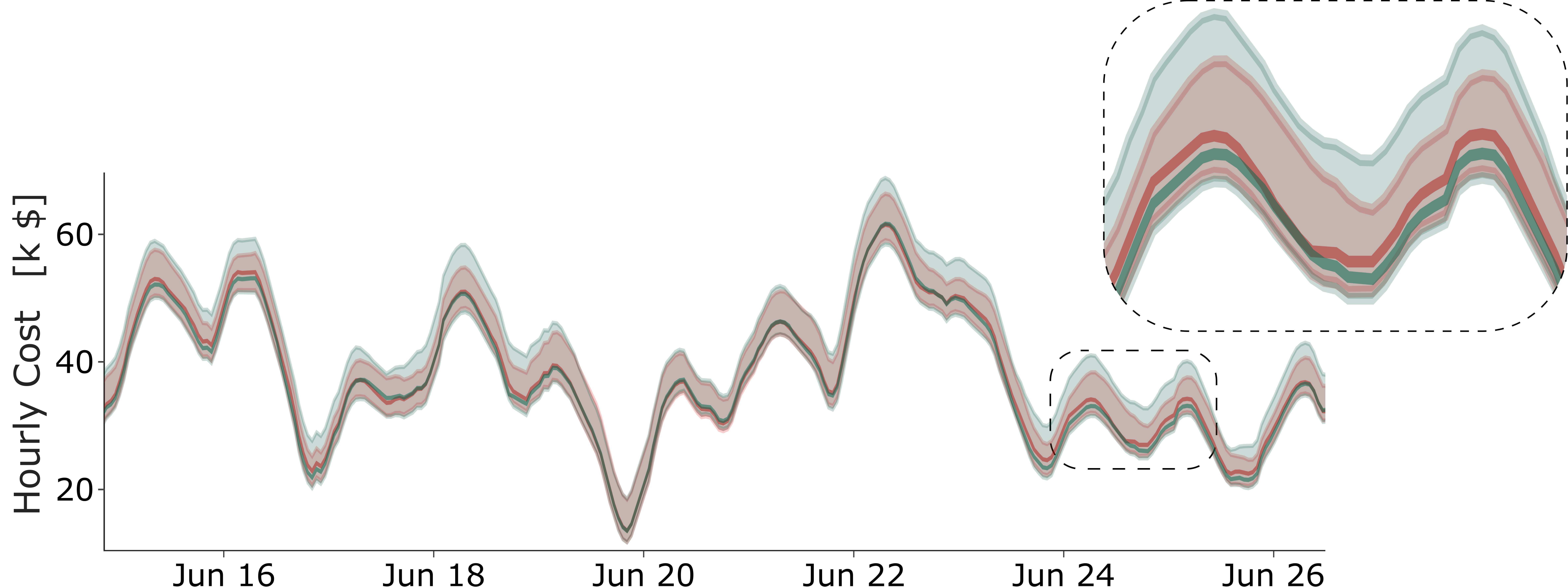}
  \caption{
  Range of hourly cost for congestion management for simulated real-time forecast errors $\bm{\omega}_t$.
  The green line shows the average congestion-management cost of the \textit{FBMC$^+$} case and the green area around it shows the corresponding cost range across all scenarios.
  The red line shows the average congestion-management cost of the \textit{FBMC$^+$ CC} case and the red area around it shows the corresponding cost range across all scenarios.}
  \label{fig:nrel-redispatch_with_omega}
\end{figure}

Fig.~\ref{fig:nrel-redispatch_with_omega} visualizes the range of hourly cost for congestion management for randomized different real-time deviations. The blue band shows the range of cost for \textit{FBMC$^+$} and the red band show cost for \textit{FBMC$^+$ CC} that includes the \glspl{frm}. The solid lines are the hourly cost without real-time deviations. 
The figure shows that, while often aligned, the \textit{FBMC$^+$ CC} case provides a tighter band that on average provides lower cost. 

\subsection{Scalability}
\label{ssec:scalability}
The proposed computation of probabilistic FRMs using chance-constraints is scalable given the mathematical properties of the resulting convex second-order conic program in \eqref{eqs:cc_economic_dispspatch} \cite{bienstock_chanceconstrained_2014} and the fact that the number of chance-constraints primarily depends on the number of zones and \glspl{cnec}.
To demonstrate this, we have applied the proposed approach to a model of the European \gls{cwe} region based on the data presented in \cite{weinhold_evaluating_2021}.
This model considers 5 zones with 1650 network nodes, 5760 generators (2022 of which are intermittent renewable generators), and 3276 transmission lines (of which an average of 141 per timestep have been identified as \glspl{cnec} using the method described in \cite{weinhold_fast_2020}).
For a \unit[24]{h} model horizon, the model was solved with the same hardware as discussed in Section~\ref{ssec:nrel-fb_parameters} above within \unit[967]{s} ($\approx$16 minutes).
The resulting cost improvement was similar to the results observed in this paper and ranged between $0.8\%$ and $1.5\%$ depending on renewable power injections. 
A more detailed analysis and discussion of a renewable-dominant European case study along the lines of \cite{weinhold_evaluating_2021} is beyond the scope of this paper and will be the subject of future work. 

\section{Conclusion}
This paper discusses the effectiveness of flow-based market coupling (FBMC), the target allocation process for cross-border trading capacity in the European electricity market, in the context of a renewable-dominant power system.
First, we derived a detailed \gls{fbmc} modeling and simulation approach, which we used to highlight the role of potential policy-specific adjustments introduced by regulatory agencies. 

Second, we discuss the effectiveness of \gls{fbmc} in the presence of uncertainty in the look-ahead ($D$--2, and $D$--1) system forecasts. 
We described that these errors are considered by \gls{tso}'s with risk-based but \textit{static} security margins on critical transmission equipment and current academic \gls{fbmc} models approximate this approach with fixed margins as a percentage of the transmission line capacity.
Motivated by these shortcomings, we proposed a risk-based method that \textit{dynamically} allocates security margins depending on forecast error statistics and a linear control policy that resembles generation shift keys (an established concept of \gls{fbmc}, thus making our method accessible for \gls{fbmc} stakeholders).
The proposed formulation leverages chance constraints and is tractable for large systems and a suitable extension to modeling \gls{fbmc}.

In our numerical experiments on the common IEEE 118 bus test system, we investigated the effectiveness of \gls{fbmc} for various renewable generation scenarios in reference to net transfer capacities (\glspl{ntc}, a reference process with static cross-border trading) and a nodal market (a theoretical optimal benchmark).
While, as expected, \gls{fbmc} can not achieve cost as low as the nodal reference, it systematically outperforms static \glspl{ntc} but performs worse in the more regulated configuration.
We demonstrated the effectiveness of the proposed chance-constrained approach, showing a significant systematic cost reduction.
Lastly, we demonstrated the scalability of the proposed probabilistic \gls{frm} computation to system models of realistic size. 

\section*{Acknowledgements}
The authors gratefully acknowledge the support by the German Federal Ministry for Economic Affairs and Energy (BMWi) in the project MODEZEEN (03EI1019B).

\bibliographystyle{IEEEtran}
\bibliography{references_bibtex.bib}

\end{document}